\documentclass[aps,pra,twocolumn,groupedaddress,superscriptaddress,nofootinbib,notitlepage,floatfix]{revtex4-2}
\usepackage{times,bm,bbm,amssymb,amsmath,amsfonts,graphics,graphicx,color,hyperref}
\usepackage[usenames,dvipsnames]{xcolor}
\hypersetup{colorlinks,linkcolor={blue},citecolor={blue},urlcolor= {blue}}
\newcommand{\ignore}[1]{}
\DeclareFontFamily{OT1}{pzc}{}
\DeclareFontShape{OT1}{pzc}{m}{it}
              {<-> s * [1.25] pzcmi7t}{}
\DeclareMathAlphabet{\mathpzc}{OT1}{pzc}
                                 {m}{it}
\usepackage[T1]{fontenc}
\usepackage{newtxtext,newtxmath}
\begin{document}

\title{Unfolding system-environment correlation in open quantum systems: Revisiting master equations and the Born approximation}

\author{A. P. Babu}
\email{aravindpbabu78@gmail.com}
\affiliation{MSP Group in QTF Center of Excellence, Department of Applied Physics, Aalto University, P. O. Box 11000, FI-00076 Aalto, Espoo, Finland}
\affiliation{Nano and Molecular Systems Research Unit, University of Oulu, P. O. Box 3000, FI-90014 Oulu, Finland}

\author{S. Alipour}
\email{s.alipoor@gmail.com}
\affiliation{MSP Group in QTF Center of Excellence, Department of Applied Physics, Aalto University, P. O. Box 11000, FI-00076 Aalto, Espoo, Finland}

\author{A. T. Rezakhani}
\affiliation{Department of Physics, Sharif University of Technology, Tehran 14588, Iran}

\author{T. Ala-Nissila}
\affiliation{MSP Group in QTF Center of Excellence, Department of Applied Physics, Aalto University, P. O. Box 11000, FI-00076 Aalto, Espoo, Finland}
\affiliation{Interdisciplinary Centre for Mathematical Modelling and Department of Mathematical Sciences, Loughborough University, Loughborough, Leicestershire LE11 3TU, U.K.}

%%%%%%%%%%%%%%%%%%%%%%%%%%%%%%%%%%%%%%%%%%%%%%%%%%%%%%%%%%%%%%%%
\begin{abstract}  
Understanding system-environment correlations in open quantum systems is vital for various quantum information and technology applications. However, these correlations are often overlooked or hidden in derivations of open-quantum-system master equations, especially when applying the Born approximation. To address this issue, given a microscopic model, we demonstrate how to retain system-environment correlation within commonly used master equations, such as the Markovian Lindblad, Redfield, second-order time-convolutionless, second-order Nakajima-Zwanzig, and second-order universal Lindblad-like equations. We show that each master equation corresponds to a particular approximation on the system-environment correlation operator. In particular, our analysis exposes the form of the hidden system-environment correlation in the Markovian Lindblad equation derived using the Born approximation. We also identify that the processes leading to the Redfield equation yield an inaccurate initial-time system-environment correlation approximation. By fixing this problem, we propose a corrected Redfield equation with an improved prediction for early stages of the time evolution. We further illustrate our results in two examples, which imply that the second-order universal Lindblad-like equation captures correlation more accurately than the other standard master equations.
\end{abstract}
\date{\today}
\maketitle

%%%%%%%%%%%%%%%%%%%%%%%%%%%%%%%%%%%%%%%%%%%%%%%%%%%%%%%%%%%%%%%%
\section{Introduction}
\label{sec:1}

Correlation in quantum systems is both a central concept in quantum information science and a vital resource for various quantum technology applications \cite{modi, horodeckis, correlationreview, q.discord, allq, vedral-1, vedral-2, CorrelationsforcomputationNPJ}, such as teleportation, dense coding, cryptography, tomography, metrology, simulation, and computation \cite{CorrelationsforcomputationNPJ, quantuminternetxximble, cryptographyreview, Quantumdiscordandcryptography, q.tomography1, q.tomography1-, q.tomography1--, q.tomography1---, q.tomography2, q.tomography3, q.tomography4, Quantummetrology0, Quantummetrology, Quantummetrology2, metrology-huelga, metrology-escher, metrology-us, RDD, q.simulation, q.computation}. In addition to technological applications, correlation provides new perspectives and methods for dealing with various fundamental problems in quantum physics. Studying quantum phase transitions \cite{QPT, Wu-Lidar, Alipour-Karimipour}, explaining the emergence of time in quantum mechanics \cite{q.time-1, q.time-2}, understanding thermalization of quantum systems \cite{thermalizationJPiilo}, and mechanisms of energy exchange between two quantum systems \cite{q.thermo-us, q.thermo-ss,q.thermo-rivas, tomas, Kawai} are examples of such problems. 
  
Dealing with most of these problems involves studying dynamics of open quantum systems, where keeping track of system-environment correlations during the course of dynamics is crucial. These correlations have diverse nature---they can be purely classical, quantum (with different forms such as entanglement and discord), or both \cite{modi,correlationreview}. Given such a diversity of correlations, discerning roles of each type of correlation in dynamics or other properties of an open system is not straightforward. Nevertheless, one can still represent system-environment correlations faithfully with a single correlation operator $\chi$, defined as the difference between the total system-environment state $\varrho_{\mathsf{SB}}$ and the uncorrelated product of the states of the system $\varrho_{\mathsf{S}}$ and the environment $\varrho_{\mathsf{B}}$, i.e,
\begin{align}
\chi=\varrho_{\mathsf{SB}} - \varrho_{\mathsf{S}} \otimes \varrho_{\mathsf{B}}.
\label{correlation}
\end{align} 
This operator encompasses all types of correlations and is thus the most relevant quantity to investigate as the ``correlation.'' It has already been shown that this operator plays an essential role in dynamical properties of open systems \cite{PRX, Ignatyuk_Morozov}. 

Much progress in the theory of open quantum systems has been made and various approximate techniques have been developed. However, in most of the existing techniques, approximations fold or mask the information of system-environment correlations into the master equation of the system.  As a result, reduced system master equations are basically meant to provide dynamical properties of the system only. For example, in the standard weak-coupling Lindblad and Redfield master equations, the explicit dependence of the dynamics on the correlation is discarded in the derivation with the Born approximation. This approximation projects the correlated state of the total system to the uncorrelated product state of the system and the environment \cite{book:Breuer-Petruccione, book:Rivas-Huelga}. However, further investigations have shown that correlation can exist between the system and environment even when the Born approximation has been applied \cite{Rivas-damped, Brask}. This implies that correlation is somehow masked even in Markovian master equations. In other techniques, such as the second-order Nakajima-Zwanzig equation (NZ$2$) and the second-order time-convolutionless master equations (TCL$2$), the approximation is based on expanding the exact dynamics in terms of the system-environment coupling \cite{book:Breuer-Petruccione, book:Rivas-Huelga, Nakajima, Zwanzig, Breuer-proj, devega, Maniscalco, Breuer-Kappler}. Although in NZ$2$ and TCL$2$ there are no strong approximations such as the Born one, correlation becomes inconspicuous in the resulting master equations. Despite many efforts to take into account correlation beyond traditional approximations \cite{Nazir-etal, pollock-1, paz-silva}, the exact form and role of system-environment correlation in open-system master equations have remained mostly elusive. Recently, by introducing a universal Lindblad-like (ULL) dynamical master equation \cite{PRX} it has been possible to treat the system-environment correlation operator explicitly, which enables approximations of correlation to different orders of the interaction Hamiltonian. Retaining correlation to first order gives the ULL$2$ master equation (or a Markovian reduction thereof, referred to as MLL). However, an important open question persists: what is the explicit form of the approximate correlation in the commonly used master equations and how does it affect their domain of validity? Note that this question goes beyond investigating how \textit{initial} correlations may affect the dynamics \cite{in-cor-1, in-cor-2, in-cor-3, in-cor-4, in-cor-5, in-cor-6, in-cor-7, in-cor-8}.

Here, we address this fundamental question and show that there is more to the master equations: we can recast their derivations such that---within a given microscopic model for the system and the environment---the role of correlation and how it is approximated become unfolded. This offers an alternative approach to the derivation of master equations and the role of the Born approximation therein. It also enables one to monitor how an approximation of the system-environment correlation (as embedded in each reduced master equation) develops when the system evolves. We emphasise that here we do not describe how the correlation should be approximated to obtain an accurate master equation (as done in Ref. \cite{PRX}). Rather, we shall focus on the information and mathematical form of the system-environment correlation contained in many standard master equations that have been widely used in the literature.

%%%%%%%%%%%%%%%%%%%%%%%%%%%%%%%%%%%%%%%%%%%%%%%%%%%%%%%%%%%%%%%%
\begingroup
\squeezetable
\begin{table*}[tp]
\caption{Approximate correlation operator $\boldsymbol{\chi}(\tau)$ in {various} master equations. All results have been written by assuming that the initial system-environment state is uncorrelated ($\boldsymbol{\chi}(0)=0$) and that the initial state of the environment satisfies the condition $\mathrm{Tr}_{\mathsf{B}}[\boldsymbol{H}_{\mathrm{I}}(\tau) \boldsymbol{\varrho}_{\mathsf{B}}(0)]=0$.}
\label{table}
\begin{ruledtabular}
\begin{tabular}{l c}
Technique & Correlation operator $\boldsymbol{\chi}(\tau)$ \\
\hline
NZ$2$ & $-i\textstyle{\int_{0}^{\tau}}  ds \, \big([\boldsymbol{H}_{\mathrm{I}}(s),\boldsymbol{\varrho}_{\mathsf{S}}(s)\otimes \boldsymbol{\varrho}_{\mathsf{B}}(0)]- {\boldsymbol{\varrho}_{\mathsf{S}}(0)} \otimes \mathrm{Tr}_{\mathsf{S}}[\boldsymbol{H}_{\mathrm{I}}(s),\boldsymbol{\varrho}_{\mathsf{S}}(s)\otimes \boldsymbol{\varrho}_{\mathsf{B}}(0)]]\big)$ \\
TCL$2$ &$-i\textstyle{\int_{0}^{\tau}}  ds \, \big([\boldsymbol{H}_{\mathrm{I}}(s),\boldsymbol{\varrho}_{\mathsf{S}}(\tau) \otimes \boldsymbol{\varrho}_{\mathsf{B}}(0)]-{\boldsymbol{\varrho}_{\mathsf{S}}(0)} \otimes \mathrm{Tr}_{\mathsf{S}}[\boldsymbol{H}_{\mathrm{I}}(s),\boldsymbol{\varrho}_{\mathsf{S}}(\tau)\otimes \boldsymbol{\varrho}_{\mathsf{B}}(0)]\big)$ \\
Redfield & $-i\textstyle{\int_{0}^{\infty}}  ds \, \big([\boldsymbol{H}_{\mathrm{I}}(\tau-s),\boldsymbol{\varrho}_{\mathsf{S}}(\tau)\otimes \boldsymbol{\varrho}_{\mathsf{B}}(0)]-{\boldsymbol{\varrho}_{\mathsf{S}}(0)} \otimes\mathrm{Tr}_{\mathsf{S}}[\boldsymbol{H}_{\mathrm{I}}(\tau-s),\boldsymbol{\varrho}_{\mathsf{S}}(\tau)\otimes \boldsymbol{\varrho}_{\mathsf{B}}(0)]\big)=:\boldsymbol{\chi}^{\mathrm{R}}(\tau)$ \\
Corrected Redfield &  $\boldsymbol{\chi}^{\mathrm{R}}(\tau)-\boldsymbol{\chi}^{\mathrm{R}}(0)$ \\ 
Lindblad &$\big\{\chi_{\omega}(\tau)=-i\textstyle{\int_{0}^{\infty}} ds\, \big([\boldsymbol{H}_{\mathrm{I}}(-\omega;\tau-s),\boldsymbol{\varrho}_{\mathsf{S}}(\tau) \otimes \boldsymbol{\varrho}_{\mathsf{B}}(0)]-{ \boldsymbol{\varrho}_{\mathsf{S}}(0)} \otimes\mathrm{Tr}_{\mathsf{S}}[\boldsymbol{H}_{\mathrm{I}}(-\omega;\tau-s),\boldsymbol{\varrho}_{\mathsf{S}}(\tau) \otimes \boldsymbol{\varrho}_{\mathsf{B}}(0)]\big)\big \}_{\omega \in \Omega}$\\
ULL$2$ &$-i \textstyle{\int_{0}^{\tau}}  ds \, [\widetilde{\boldsymbol{H}}_{\mathrm{I}}(s),\boldsymbol{\varrho}_{\mathsf{S}}(s)\otimes \boldsymbol{\varrho}_{\mathsf{B}}(s)]$ \\ 
MLL & $-i \tau [\widetilde{\boldsymbol{H}}_{\mathrm{I}}(\tau),\boldsymbol{\varrho}_{\mathsf{S}}(\tau)\otimes \boldsymbol{\varrho}_{\mathsf{B}}(\tau)]$ \\
\end{tabular}
\end{ruledtabular}
\end{table*}
\endgroup
%%%%%%%%%%%%%%%%%%%%%%%%%%%%%%%%%%%%%%%%%%%%%%%%%%%%%%%%%%%%%%%%

The key observation is rewriting an exact dynamical equation of the reduced system $\mathsf{S}$ in a form in which correlation appears explicitly. Comparing each master equation obtained from approximate techniques with the exact dynamics---where correlation is explicit---we unravel how the system-environment correlation operator has been approximated in the Lindblad, Redfield, TCL$2$, and NZ$2$ techniques. We then compare accuracy of these approximate correlations in an example. We demonstrate that the ULL$2$ correlation is closer to the exact result as compared to the other techniques here. Furthermore, deriving correlation in the Redfield equation reveals that its dynamics is afflicted with an inconsistency in the initial system-environment correlation which makes it nonzero despite that the initial condition implies otherwise. We then introduce a corrected version of the Redfield equation by resolving this problem and show that the new equation captures the initial dynamics better than Redfield. In this respect, our approach may be complementary to the version of the Redfield equation with ``slippage'' of initial conditions \cite{slippage}. 

%%%%%%%%%%%%%%%%%%%%%%%%%%%%%%%%%%%%%%%%%%%%%%%%%%%%%%%%%%%%%%%%
\section{Correlation in open-quantum system dynamics}
\label{sec:2}

We assume that {we have a known microscopic model in which the system (with the Hamiltonian $H_{\mathsf{S}}$) interacts with a given environment (with the Hamiltonian $H_{\mathsf{B}}$), which is initially in a known state $\varrho_{\mathsf{B}}(0)$, through the interaction Hamiltonian $H_{\mathrm{I}}$, such that for the total system we have $H_{\mathsf{SB}}=H_{\mathsf{S}}+H_{\mathsf{B}}+H_{\mathrm{I}}$}. To see how correlation appears in the master equation of the system, we start from the Schr\"{o}dinger equation of the total system $\dot{\varrho}_{\mathsf{SB}}(\tau)=-i [H_{\mathsf{SB}},\varrho_{\mathsf{SB}}(\tau)]$, where dot denotes time derivative $d/d\tau$ and we work in dimensionless units where $\hbar \equiv k_{\mathsf{B}}\equiv 1$. By inserting $\varrho_{\mathsf{SB}}(\tau)$ in terms of $\chi(\tau)$, we obtain
\begin{align}
\dot{\boldsymbol{\varrho}}_{\mathsf{S}}(\tau)&=-i \big[\mathrm{Tr}_{\mathsf{B}}[\boldsymbol{H}_{\mathrm{I}}(\tau)\boldsymbol{\varrho}_{\mathsf{B}}(\tau)],\boldsymbol{\varrho}_{\mathsf{S}}(\tau)\big]-i \mathrm{Tr}_{\mathsf{B}}[\boldsymbol{H}_{\mathrm{I}}(\tau),\boldsymbol{\chi}(\tau)],
\label{exactS-intpic}
\end{align}
where the boldface notation $\boldsymbol{O} \equiv e^{i (H_{\mathsf{S}}+H_{\mathsf{B}})\tau} O \,e^{-i (H_{\mathsf{S}}+H_{\mathsf{B}})\tau}$ denotes the interaction picture for any arbitrary operator $O$. See Appendix \ref{app:exactD} for derivation of Eq. \eqref{exactS-intpic}.

In principle, each technique for deriving a master equation (with its own set of assumptions or approximations) constructs a basic form of a dynamical equation in which the elements of the microscopic model are present and discernible. The universality and exactness of Eq. \eqref{exactS-intpic} indicate that master equations can also be reformatted and brought into a form almost similar to this exact form. Thus, by comparing the reformatted forms with the right-hand side of Eq. \eqref{exactS-intpic} we identify the factors appearing in the second term $\mathrm{Tr}_{\mathsf{B}}[\boldsymbol{H}_{\mathrm{I}}(\tau),\cdot]$ and thereby read off some specific \textit{ansatzes} for approximate $\boldsymbol{\chi}(\tau)$s---Secs. \ref{sec:nz2} -- \ref{sec:mll}. We then confirm these specific ansatzes as reasonable candidates for (approximate) correlation through an alternative approach to deriving master equations by directly applying relevant approximations on the exact correlation operator obtained from the Schr\"{o}dinger equation of the total system. This direct approach, which is delineated in Appendix \ref{app:new}, indicates that the Born approximation plays a special role in obtaining legitimate approximations of the correlation operator for master equations. Table \ref{table} summarizes the results for the approximated correlation for several standard master equations. We should also remark that, similar to the common practice in derivation of master equations, two assumptions have been applied throughout our analysis, namely that $\chi(0)=0$ and $\mathrm{Tr}_{\mathsf{B}}[\boldsymbol{H}_{\mathrm{I}}(\tau)\boldsymbol{\varrho}_{\mathsf{B}}(0)]=0$.

%%%%%%%%%%%%%%%%%%%%%%%%%%%%%%%%%%%%%%%%%%%%%%%%%%%%%%%%%%%%%%%%
\subsection{System-environment correlation in NZ2}
\label{sec:nz2} 
 
We start from the second-order approximation of the NZ equation given by \cite{book:Breuer-Petruccione,book:Rivas-Huelga,PRX} 
\begin{align}
\dot{\boldsymbol{\varrho}}_{\mathsf{S}}(\tau)=-\mathrm{Tr}_{\mathsf{B}}\big[\boldsymbol{H}_{\mathrm{I}}(\tau), \textstyle{\int_{0}^{\tau}}  ds \, [\boldsymbol{H}_{\mathrm{I}}(s),\boldsymbol{\varrho}_{\mathsf{S}}(s)\otimes \boldsymbol{\varrho}_{\mathsf{B}}(0)] \big].
\label{NZ$2$-1}
\end{align}
Comparing the NZ$2$ equation with the exact dynamics of Eq. \eqref{exactS-intpic}, it can be seen that in the first term $\boldsymbol{\varrho}_{\mathsf{B}}(\tau)$ has been approximated by $\boldsymbol{\varrho}_{\mathsf{B}}(0)${, which now by the initial environment condition cancels this term,} and from the second term one may read a candidate $\boldsymbol{X}(\tau)$ for the correlation as $\textstyle{\int_{0}^{\tau}} ds \, [\boldsymbol{H}_{\mathrm{I}}(s),\boldsymbol{\varrho}_{\mathsf{S}}(s)\otimes \boldsymbol{\varrho}_{\mathsf{B}}(0)]$. However, this {candidate} does not {yet} satisfy the vanishing partial trace condition on the system part for the correlation operator---while its partial trace on the environment part is zero due to the environment initial condition. To remedy this problem, we replace $\boldsymbol{X}$ with the modified form $\boldsymbol{\chi}=\boldsymbol{X}-\boldsymbol{A}\otimes \mathrm{Tr}_{\mathsf{S}}[\boldsymbol{X}]$, for some Hermitian $\boldsymbol{A}$. The condition $\mathrm{Tr}_{\mathsf{B}}[\boldsymbol{\chi}]=0$ is now met, for any $\boldsymbol{A}$, simply by the initial environment condition and that $\boldsymbol{X}$ is a commutator. The condition $\mathrm{Tr}_{\mathsf{S}}[\boldsymbol{\chi}]=0$ requires that $\mathrm{Tr}_{\mathsf{S}}[\boldsymbol{A}]=1$. We note that there is no unique choice for $\boldsymbol{A}$, but a natural one is $\boldsymbol{A}=\boldsymbol{\varrho}_{\mathsf{S}}(0)$. Thus, we propose the ansatz
\begin{align}
\boldsymbol{\chi}^{\textsc{nz}{\scriptscriptstyle{2}}}(\tau)=&-i\textstyle{\int_{0}^{\tau}} ds \, \big([\boldsymbol{H}_{\mathrm{I}}(s),\boldsymbol{\varrho}_{\mathsf{S}}(s)\otimes \boldsymbol{\varrho}_{\mathsf{B}}(0)]\label{corr-NZ$2$} \nonumber\\
&- \boldsymbol{\varrho}_{\mathsf{S}}(0) \otimes\mathrm{Tr}_{\mathsf{S}}[\boldsymbol{H}_{\mathrm{I}}(s),\boldsymbol{\varrho}_{\mathsf{S}}(s)\otimes \boldsymbol{\varrho}_{\mathsf{B}}(0)] \big).
\end{align}
Note that the extra term in $\boldsymbol{\chi}^{\textsc{nz}{\scriptscriptstyle{2}}}(\tau)$ has no impact on the system dynamics because of the cyclicity of $\mathrm{Tr}_{\mathsf{B}}$ over the Hilbert space of the environment. Comparing this form with the direct calculations of Appendix \ref{app:new} verifies that this $\boldsymbol{\chi}^{\textsc{nz}{\scriptscriptstyle{2}}}(\tau)$ is indeed correct.
 
%%%%%%%%%%%%%%%%%%%%%%%%%%%%%%%%%%%%%%%%%%%%%%%%%%%%%%%%%%%%%%%%
\subsection{System-environment correlation in TCL$2$}
\label{sec:3}

The TCL$2$ dynamical equation is the time-local version of the NZ$2$ equation, and it is obtained by changing $\boldsymbol{\varrho}_{\mathsf{S}}(s)$ in the integrand of Eq. \eqref{NZ$2$-1} to $\bf{\varrho}_{\mathsf{S}}(\tau)$ \cite{book:Breuer-Petruccione,book:Rivas-Huelga},
\begin{equation}
\dot{\boldsymbol{\varrho}}_{\mathsf{S}}(\tau) =-\mathrm{Tr}_{\mathsf{B}}\big[\boldsymbol{H}_{\mathrm{I}}(\tau),\textstyle{\int_{0}^{\tau}} ds\,  [\boldsymbol{H}_{\mathrm{I}}(s),\boldsymbol{\varrho}_{\mathsf{S}}(\tau) \otimes \boldsymbol{\varrho}_{\mathsf{B}}(0)]\big],
\label{TCL$2$-1}
\end{equation}
which again has been obtained under the environment initial condition assumption. It is evident then that the above $\boldsymbol{\chi}^{{\textsc{tcl}{\scriptscriptstyle{2}}}}(\tau)$ can also be obtained by applying the same change on $\boldsymbol{\chi}^{{\textsc{nz}{\scriptscriptstyle{2}}}}(\tau)$. Similarly to the NZ$2$ case, the added term in $\boldsymbol{\chi}^{\textsc{tcl}{\scriptscriptstyle{2}}}(\tau)$ does not change the dynamical equation of the system. In addition, Appendix \ref{app:new} verifies that this is the correct correlation.

%%%%%%%%%%%%%%%%%%%%%%%%%%%%%%%%%%%%%%%%%%%%%%%%%%%%%%%%%%%%%%%%
\subsection{System-environment correlation in Redfield}
\label{sec:redfield} 
 
The Redfield {(R)} equation under the environment initial condition assumption and before the Markov approximation is equivalent to TCL$2$ \cite{book:Breuer-Petruccione,book:Rivas-Huelga}. However, after applying the Markov approximation---replacing the upper limit of the integral in Eq. \eqref{TCL$2$-1} with $\infty$ and changing $s\to \tau-s$ in $\boldsymbol{H}_{\mathrm{I}}(s)$---this gives 
\begin{equation}
\dot{\boldsymbol{\varrho}}_{\mathsf{S}}(\tau) =-\mathrm{Tr}_{\mathsf{B}}\big[\boldsymbol{H}_{\mathrm{I}}(\tau),\textstyle{\int_{0}^{\infty}} ds\, [\boldsymbol{H}_{\mathrm{I}}(\tau-s),\boldsymbol{\varrho}_{\mathsf{S}}(\tau) \otimes \boldsymbol{\varrho}_{\mathsf{B}}(0)]\big],
\label{eq:RF}
\end{equation}
which often is referred to as the time-independent Redfield equation \cite{RF=TCL, Nafari, Thingna, Becker} (for a rigorous analysis of the Redfield equation and alternative derivation of the Lindblad equation from it, see Ref. \cite{rudner}). Reading a candidate correlation $\boldsymbol{X}(\tau)$ yields an issue similar to the one in the NZ$2$ case; that is, the partial trace of this operator on the environment does not vanish. Following a similar recipe as in NZ$2$ to remove this (which is also justified in Appendix \ref{app:new}), we read the ansatz
\begin{align}
\boldsymbol{\chi}^{\textsc{r}}(\tau)=&-i\textstyle{\int_{0}^{\infty}}  ds \, \big([\boldsymbol{H}_{\mathrm{I}}(\tau-s),\boldsymbol{\varrho}_{\mathsf{S}}(\tau)\otimes \boldsymbol{\varrho}_{\mathsf{B}}(0)]\label{corr-RF}\\
& - \boldsymbol{\varrho}_{\mathsf{S}}(\tau) \otimes\mathrm{Tr}_{\mathsf{S}}[\boldsymbol{H}_{\mathrm{I}}(\tau-s),\boldsymbol{\varrho}_{\mathsf{S}}(\tau)\otimes \boldsymbol{\varrho}_{\mathsf{B}}(0)]\big). \nonumber
\end{align}
Replacing this $\boldsymbol{\chi}^{\textsc{r}}(\tau)$ in the Redfield equation does not modify the system dynamics. Correctness of this particular ansatz for $\boldsymbol{\chi}^{\textsc{r}}(\tau)$ is again verified by comparing it with the approximated form of correlation obtained directly from the Schr\"{o}dinger equation of the total system---Appendix \ref{app:new}.

Similar to the other techniques, the Redfield equation has been obtained by assuming that the initial system-environment state is uncorrelated. However, it can be seen from Eq. \eqref{corr-RF} that $\boldsymbol{\chi}^{\textsc{r}}(0)$ is not necessarily zero. This discrepancy {is the manifestation of the fact that time-independent Redfield equation cannot capture the short-time dynamics correctly} \cite{slippage, Whitney, Hartmann}. An elementary way to alleviate this deficiency is to redefine the correlation as $\boldsymbol{\chi}^{\textsc{cr}}(\tau)=\boldsymbol{\chi}^{\textsc{r}}(\tau)-\boldsymbol{\chi}^{\textsc{r}}(0)$, which vanishes at the initial time. However, subtracting $\boldsymbol{\chi}^{\textsc{r}}(0)$ leads to a modification of the {time-independent} Redfield equation and we call this the corrected Redfield (CR) equation, given by $\dot{\boldsymbol{\varrho}}_{\mathsf{S}}(\tau)=-i\mathrm{Tr}_{\mathsf{B}}[\boldsymbol{H}_{\mathrm{I}}(\tau), \boldsymbol{\chi}^{\textsc{cr}}(\tau)]$. We show later {through examples} that the CR equation indeed represents the short-time behavior of the dynamics more accurately. The CR approach can be interpreted as a ``slippage'' of the initial \textit{correlation} (rather than the initial \textit{state} of the system) and modifying the master equation in order to fix the slippage problem of the Redfield equation \cite{slippage, Whitney}.

%%%%%%%%%%%%%%%%%%%%%%%%%%%%%%%%%%%%%%%%%%%%%%%%%%%%%%%%%%%%%%%%
\subsection{System-environment correlation in Lindblad}
\label{sec:4}

By applying the rotating-wave approximation (RWA) on the Redfield equation we end up with the Lindblad {(L)} master equation \cite{book:Breuer-Petruccione}. However, since the nature of the RWA is such that it mixes the elements in the nested commutators in the correlation part of the Redfield equation, it becomes impossible to read correlation by comparing the final form of the Lindblad equation and the exact dynamics. To overcome this issue, we need to delve into the details of the derivation of the Lindblad equation. To apply the RWA, we note that any operator $O(\tau)$ (with support on the system-environment Hilbert space) can be decomposed in the eigenprojector basis $\{\Pi_E\}$ of the system Hamiltonian as $O(\tau)=\textstyle{\sum_{\omega \in \Omega}} O(\omega;\tau)$ such that $O(\omega;\tau) \equiv \sum_{E',E|E'-E=\omega}\Pi_{E}\otimes \mathbbmss{I}_{\textsc{B}}\, O(\tau)\,\Pi_{E'} \otimes \mathbbmss{I}_{\textsc{B}}$, where $\Omega\equiv\{E-E' | E,E' \,\text{are eigenvalues of} \, H_{\mathsf{S}}\}$ is the set of the energy gaps of the system Hamiltonian. It should be noted that $O(\omega;\tau)$ is not necessarily Hermitian; in fact, $O^{\dag}(\omega;\tau)=O(-\omega;\tau)$. Applying this decomposition on the interaction Hamiltonian $H_{\mathrm{I}}=\sum_{k} S_{k} \otimes B_{k}$, we obtain in the interaction picture $\boldsymbol{H}_{\mathrm{I}}(\tau)=\sum_{\omega\in\Omega}\boldsymbol{H}_{\mathrm{I}}(\omega;\tau)$, where $\boldsymbol{H}_{\mathrm{I}}(\omega;\tau)=\textstyle{\sum_{k}} e^{-i\omega \tau} S_{k}(\omega)\otimes \boldsymbol{B}_{k}(\tau)$. Replacing this expansion into Eq. \eqref{eq:RF}, and applying the RWA which eliminates the terms with $\omega+\omega'\neq 0$, we get   
\begin{align}
&\dot{\boldsymbol{\varrho}}_{\mathsf{S}}(\tau) =\label{eq:LB}\\
&-\textstyle{\sum_{\omega}} \mathrm{Tr}_{\mathsf{B}}\big[\boldsymbol{H}_{\mathrm{I}}(\omega;\tau),\textstyle{\int_{0}^{\infty}} ds\, [\boldsymbol{H}_{\mathrm{I}}(-\omega;\tau-s),\boldsymbol{\varrho}_{\mathsf{S}}(\tau) \otimes \boldsymbol{\varrho}_{\mathsf{B}}(0)]\big].
\nonumber
\end{align}  
%%%%%%%%%%%%%%%%%%%%%%%%%%%%%%%%%%%%%%%%%%%%%%%%%%%%%%%%%%%%%%%%
\begin{figure*}[tp]
\includegraphics[width=.85\linewidth]{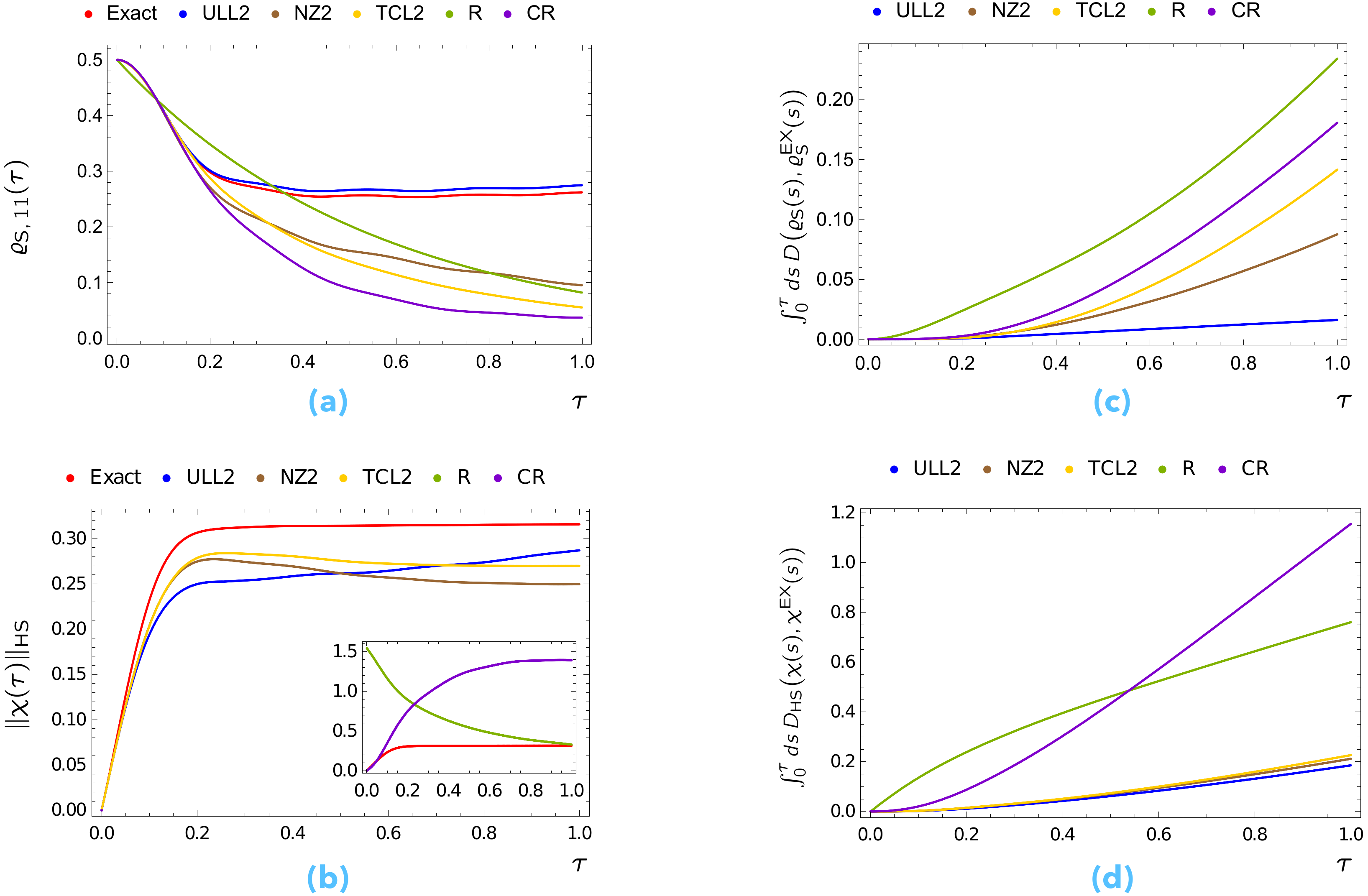}
\caption{Qubit in an Ohmic environment of qubits. (\textbf{\textsf{a}}) Dynamics of the excited-state population $\varrho_{\mathsf{S}, 11}$ and (\textbf{\textsf{b}}) norm of the correlation operator in the exact dynamics (``Exact'' \& ``EX''), ULL$2$, NZ$2$, TCL$2$, Redfield (``R'') which is equivalent to the Lindblad, and corrected Redfield (``CR'') techniques. (\textbf{\textsf{c}}) Errors in capturing the dynamics of the system state and (\textbf{\textsf{d}}) the correlation {\it vs.} time. Here $M=255$, $\omega_c/\omega_{0}=10$, $\eta/\omega_{0}=1$ (setting $\omega_{0}=1$), $\Delta \omega=0.1$, $\delta \tau=0.0005$, and the initial state of the system is $(|0\rangle+|1\rangle)/\sqrt{2}$. All quantities are in natural units where $\hbar\equiv k_{\mathrm{B}} \equiv 1$.}
\label{Ohmic1}
\end{figure*}
%%%%%%%%%%%%%%%%%%%%%%%%%%%%%%%%%%%%%%%%%%%%%%%%%%%%%%%%%%%%%%%%
This is the Lindblad equation. To recast it in the well-known Lindblad form one only needs to expand each term and calculate the partial trace \cite{book:Breuer-Petruccione}. Defining (the mode or frequency component)
\begin{align}
\chi_{\omega}(\tau)\equiv&-i\textstyle{\int_{0}^{\infty}} ds\big([\boldsymbol{H}_{\mathrm{I}}(-\omega;\tau-s),\boldsymbol{\varrho}_{\mathsf{S}}(\tau) \otimes \boldsymbol{\varrho}_{\mathsf{B}}(0)]\nonumber\\
&-\boldsymbol{\varrho}_{\mathsf{S}}(0) \otimes\mathrm{Tr}_{\mathsf{S}}[\boldsymbol{H}_{\mathrm{I}}(-\omega;\tau-s),\boldsymbol{\varrho}_{\mathsf{S}}(\tau)\otimes \boldsymbol{\varrho}_{\mathsf{B}}(0)]\big),\nonumber
\end{align}
where the second term has been added to satisfy the vanishing partial trace conditions, and noting that 
\begin{equation}
\dot{\boldsymbol{\varrho}}_{\mathsf{S}}(\tau) = \textstyle{\sum_{\omega\in\Omega}} \dot{\boldsymbol{\varrho}}_{\mathsf{S}}(\omega;\tau),
\end{equation}
Eq. \eqref{eq:LB} can be decomposed into a set of coupled equations
\begin{align}
\dot{\boldsymbol{\varrho}}_{\mathsf{S}}(\omega;\tau) =-i\,\mathrm{Tr}_{\mathsf{B}}[\boldsymbol{H}_{\mathrm{I}}(\omega;\tau),\chi_{\omega}(\tau)], \quad \omega \in \Omega.
\end{align} 
From this we can conclude that the Lindblad equation treats the system-environment correlation through a set of correlations $\{\chi_{\omega}(\tau)\}_{\omega \in \Omega}$ between each energy-gap mode of the system and the environment. Note that the total Lindblad correlation $\boldsymbol{\chi}^{\textsc{l}}(\tau)\neq\sum_{\omega\in\Omega}\chi_{\omega}(\tau)$.
In fact, it is not evident (if not impossible) how to obtain a \textit{single} total system-environment correlation for the Lindblad equation. This is the complication imposed by the RWA and the very nature of the approximations in the Lindblad equation.

%%%%%%%%%%%%%%%%%%%%%%%%%%%%%%%%%%%%%%%%%%%%%%%%%%%%%%%%%%%%%%%%
\subsection{System-environment correlation in ULL$2$}
\label{sec:5}

The ULL$2$ dynamical equation has been introduced in Ref. \cite{PRX}. This equation has been obtained based on an iterative solution of the dynamics of correlation up to first order (in $\boldsymbol{H}_{\mathrm{I}}$) and it includes correlation explicitly, given by
\begin{align}
\boldsymbol{\chi}^{\textsc{ull}{\scriptscriptstyle{2}}}(\tau)=-i \textstyle{\int_{0}^{\tau}}  ds \, [\widetilde{\boldsymbol{H}}_{\mathrm{I}}(s),\boldsymbol{\varrho}_{\mathsf{S}}(s)\otimes \boldsymbol{\varrho}_{\mathsf{B}}(s)],
\label{chi-1-uncorrelated-initial}
\end{align}
where 
\begin{equation}
\widetilde{\boldsymbol{H}}_{\mathrm{I}}(\tau)=\boldsymbol{H}_{\mathrm{I}}(\tau) -\mathrm{Tr}_{\mathsf{B}}[\boldsymbol{H}_{\mathrm{I}}(\tau) \boldsymbol{\varrho}_{\mathsf{B}}(\tau)]-\mathrm{Tr}_{\mathsf{S}}[\boldsymbol{H}_{\mathrm{I}}(\tau) \boldsymbol{\varrho}_{\mathsf{S}}(\tau)].
\end{equation}
The ULL$2$ equation is given by Eq. \eqref{exactS-intpic} with $\boldsymbol{\chi}(\tau) \to \boldsymbol{\chi}^{\textsc{ull}{\scriptscriptstyle{2}}}(\tau)$. An equation similar to Eq. \eqref{exactS-intpic}, with $\mathsf{S}\leftrightarrow \mathsf{B}$, also holds for $\boldsymbol{\varrho}_{\mathsf{B}}(\tau)$ and should be included for a full description. For an independent but relatively similar technique, see also Ref. \cite{Ignatyuk_Morozov}. 

%%%%%%%%%%%%%%%%%%%%%%%%%%%%%%%%%%%%%%%%%%%%%%%%%%%%%%%%%%%%%%%%
\subsection{System-environment correlation in MLL}
\label{sec:mll}

A Markovian reduction of the ULL$2$ equation has been also worked out in Ref. \cite{PRX}, dubbed as the MLL equation,
\begin{align}
\dot{\boldsymbol{\varrho}}_{\mathsf{S}}(\tau)=&-i \big[\mathrm{Tr}_{\mathsf{B}}[\boldsymbol{H}_{\mathrm{I}}(\tau)\boldsymbol{\varrho}_{\mathsf{B}}(\tau)],\boldsymbol{\varrho}_{\mathsf{S}}(\tau) \big]\nonumber \\ & - \tau\, \mathrm{Tr}_{\mathsf{B}}\big[\boldsymbol{H}_{\mathrm{I}}{(\tau)},[\widetilde{\boldsymbol{H}}_{\mathrm{I}}(\tau),\boldsymbol{\varrho}_{\mathsf{S}}(\tau)\otimes \boldsymbol{\varrho}_{\mathsf{B}}(\tau)]\big].
\label{MLL-S}
\end{align}
This equation can be recast in the form of a Lindblad equation with positive quantum jump rates \cite{PRX}. Note that this equation approximates the ULL$2$ correlation at short times as
\begin{align}
\boldsymbol{\chi}^{\textsc{mll}}(\tau)=-i \tau [\widetilde{\boldsymbol{H}}_{\mathrm{I}}(\tau),\boldsymbol{\varrho}_{\mathsf{S}}(\tau)\otimes \boldsymbol{\varrho}_{\mathsf{B}}(0)],
\label{chi-MLLl}
\end{align}
where $\boldsymbol{\varrho}_{\mathsf{B}}(\tau)$ has been replaced by $\boldsymbol{\varrho}_{\mathsf{B}}(0)$ to remove the dependence of the system master equation on $\boldsymbol{\varrho}_{\mathsf{B}}(\tau)$. In the following we do not discuss the MLL equation as it is basically an approximation of the more accurate ULL$2$ equation.

%%%%%%%%%%%%%%%%%%%%%%%%%%%%%%%%%%%%%%%%%%%%%%%%%%%%%%%%%%%%%%%%
\section{Example: Qubit in an environment of qubits}
\label{sec:example}

We consider a qubit interacting with an environment of qubits with the total Hamiltonian
\begin{equation}
H_{\mathsf{SB}} = \omega_{0} \sigma_{+}\sigma_{-}+\textstyle{\sum_{k=1}^{M}}\omega_{k} \Sigma^{k}_{+}  \Sigma^{k}_{-} +\textstyle{\sum_{k=1}^{M}}g_{k} (\sigma_{-} \Sigma^{k}_{+}  +\sigma_{+} \Sigma^{k}_{-} ),
\label{system-H}
\end{equation}
where $\sigma_{\pm}=(\sigma_{x}\pm i\sigma_{y})/2$ are the raising and lowering operators of the system qubit, with $\sigma_x$ and $\sigma_y$ being the Pauli operators. Similarly, $\Sigma^{k}_{\pm}$ are the raising and lowering operators associated with the $k$th qubit of the environment. This model is important in the context of quantum computation applications as it can model the environment in some experimental setups for realization of a qubit \cite{NPJQI:Babu,PhysRevB:Bilmes,Review:muller}. The effect of the environment on the dynamics of the system is usually encompassed in the spectral density function $ J(\omega)=\textstyle{\sum_{k}} g_{k}^{2} \delta(\omega-\omega_{k})$. The conventional choices of spectral densities are in the continuum limit, requiring an infinite environment assumption. Assuming a given spectral density, to obtain consistent coupling constants needed for numerical simulation of the dynamics, we choose $g_{k}=\sqrt{J(\omega_{k}) \Delta \omega}$, where $\omega_{k}=k\Delta \omega $ and the parameter $\Delta \omega $ is determined by the condition $\int_{0}^{\omega_{\max}}d\omega\, J(\omega)\approx \sum_{k}^{M} g_{k}^{2} $, where $\omega_{\max}=M \Delta \omega$ \cite{Rivas-damped}. 

In the following, we study the dynamics and correlation of the model in Eq. \eqref{system-H} with the Ohmic and Lorentzian spectral density functions. We assume that the environment is initially at zero temperature, \textit{i.e.}, $\varrho_{\mathsf{B}}(0)=(|0\rangle\langle 0|)^{\otimes M}$ and $\varrho_{\mathsf{S}}(0)=|\psi\rangle\langle\psi|$, where $|\psi\rangle=(|0\rangle+|1\rangle)/\sqrt{2}$. To obtain the exact dynamics, we use the symmetry property  $[\sigma_{+} \sigma_{-} + \textstyle{\sum_{k=1}^{M}} \Sigma_{+}^{k}\Sigma_{-}^{k},H_{\mathsf{SB}}] =0$ for the initial environment state $\varrho_{\mathsf{B}}(0)=(|0\rangle\langle 0|)^{\otimes M}$. This property reduces the problem to a single-excitation subspace of the Hilbert space of the total system Hamiltonian and yields a $M+1$ subspace for the environment. Further, we have used the forward Euler discretization method \cite{Euler} to compute integrals and solve dynamical differential equations for the system and the environment with a time step $\delta \tau$ as given in the captions of the figures. It should also be noted that all types of master equations that we are studying here are of second order in the norm of the interaction Hamiltonian under a time integral. Due to this, we can estimate the time duration of the validity of these equations for predicting the open-system dynamics as $\tau \ll 1/\|H_{\rm I}\|^2$ and plot all the figures up this time for the purpose of comparison of the different master equations.

%%%%%%%%%%%%%%%%%%%%%%%%%%%%%%%%%%%%%%%%%%%%%%%%%%%%%%%%%%%%%%%%
\begin{figure*}[tp]
\includegraphics[width=.85\linewidth]{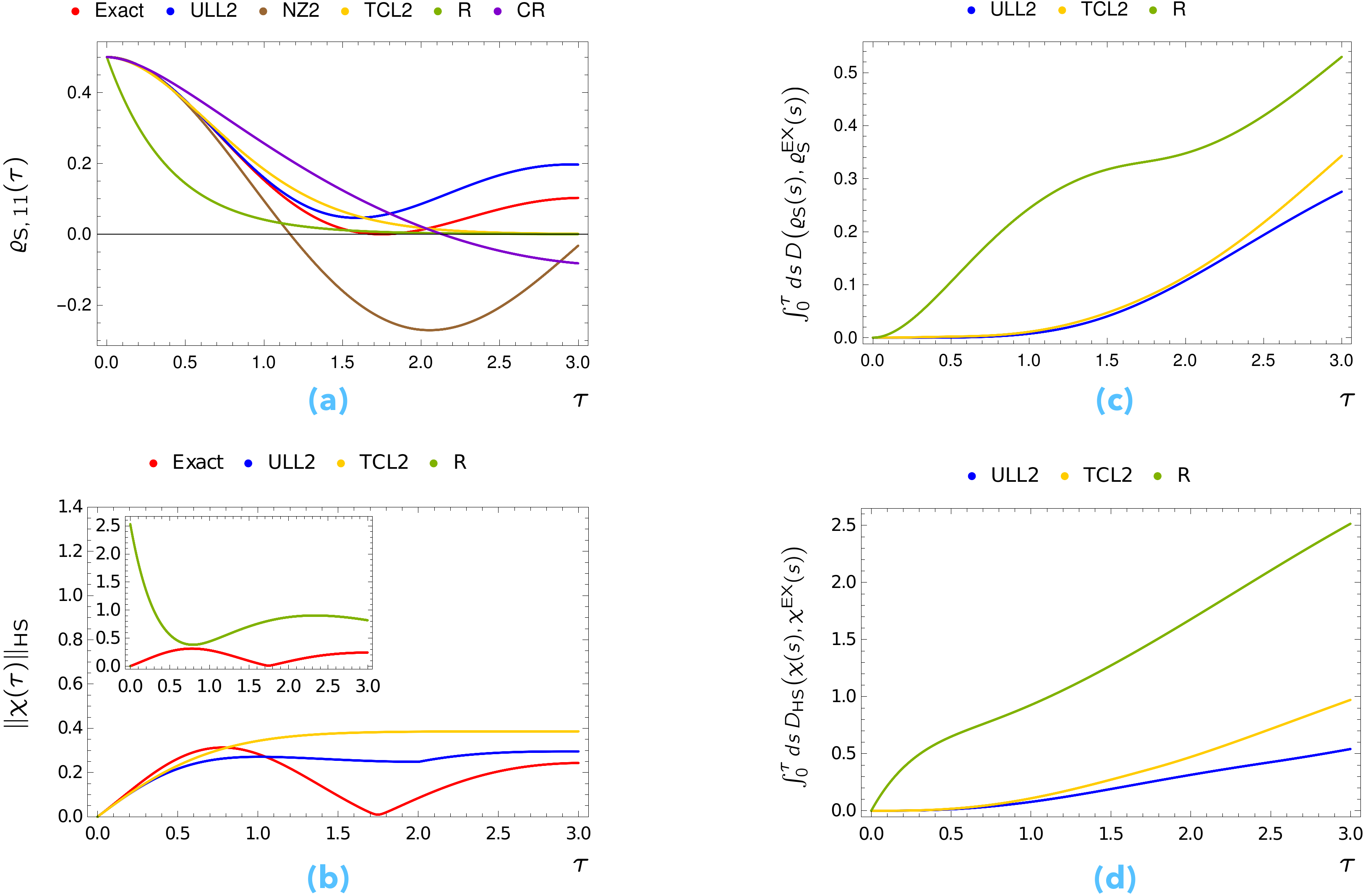}
\caption{Qubit in a Lorentzian environment of qubits. (\textbf{\textsf{a}}) Dynamics of the excited-state population $\varrho_{\mathsf{S}, 11}$, and (\textbf{\textsf{b}}) norm of the correlation operator in the exact dynamics (``Exact'' \& ``EX''), ULL$2$, NZ$2$, TCL$2$, Redfield (``R'') which is equivalent to the Lindblad, and corrected Redfield (``CR'') techniques. (\textbf{\textsf{c}}) Error in the dynamics of the system density matrix, and (\textbf{\textsf{d}}) error in capturing dynamics of the correlation \textit{vs.} time. The values of the parameters are $M=255$, $\lambda=1$, $\Gamma=5$ ($\lambda/\Gamma=0.2$), $\Delta\omega=0.05$, and $\delta \tau = 0.0005$, and the initial state of the system is $(|0\rangle+|1\rangle)/\sqrt{2}$. In this case, NZ$2$, and CR give unphysical results and thus are not plotted. All quantities are in natural units where $\hbar\equiv k_{B} \equiv 1$.}
\label{LO1}
\end{figure*}
%%%%%%%%%%%%%%%%%%%%%%%%%%%%%%%%%%%%%%%%%%%%%%%%%%%%%%%%%%%%%%%%

We have used the {Hilbert-Schmidt} norm of the correlation operator $\Vert \chi^{\scriptscriptstyle{a}}\Vert_{{\textsc{hs}}}$ to characterize  the system-environment correlation. However, it is important to note that the norm of the correlation operator cannot capture all features of correlation as it measures only one feature of an operator, \textit{i.e.}, its amplitude. It has already been discussed in Ref. \cite{PRAsahar} that scalar measures of correlation, such as mutual information and the norm of the correlation operator, cannot be relied upon as comprehensive measures of correlations.

We compare different techniques by studying their ability to capture the system-environment correlation. To this end, we use the accumulative error in correlation given by $\textstyle{\int_{0}^{\tau}} ds\, D_{\textsc{hs}}\big(\chi^{\scriptscriptstyle{a}}(s),\chi^{\textsc{ex}}(s)\big)$, where $D_{\textsc{hs}}(\chi^{\scriptscriptstyle{a}},\chi^{\textsc{ex}})= \sqrt{\mathrm{Tr}[(\chi^{\scriptscriptstyle{a}}-\chi^{\textsc{ex}})^{2} ]}$ is the Hilbert-Schmidt distance of the exact correlation $\chi^{\textsc{ex}}$ and $\chi^{\scriptscriptstyle{a}}$, with $a\in\{$ULL$2$, NZ$2$, TCL$2$, R, CR, L$\}$. To further analyze how the approximated correlation affects the accuracy of the dynamical equation in each case we also compare different techniques through error in capturing the system state defined by $\textstyle{\int_{0}^{\tau}} ds\, D \big(\varrho^{\scriptsize{a}}(s), \varrho^{\textsc{ex}}(s)\big)$, where $D (\varrho^{\scriptscriptstyle{a}}, \varrho^{\textsc{ex}}) = (1/2)\mathrm{Tr}[|\varrho^{\scriptscriptstyle{a}}-\varrho^{\textsc{ex}}|]$ is the trace distance of the density matrix in each technique $\varrho^{\scriptscriptstyle{a}}$ from the exact one $\varrho^{\textsc{ex}}$ (with $|X|=\sqrt{X^{\dag}X}$) \cite{PRAJyrkiPillo}. 

Details of this example can be found in Appendix \ref{app:example-details}. In Appendix \ref{app:JC}, we also provide another example with the Jaynes–Cummings model.

%%%%%%%%%%%%%%%%%%%%%%%%%%%%%%%%%%%%%%%%%%%%%%%%%%%%%%%%%%%%%%%%
\subsection{Ohmic environment}
\label{sec:ohmic}

Assume the Ohmic spectral density 
\begin{equation}
J(\omega)=(\eta/\pi) \omega e^{-\omega/\omega_{\mathrm{c}}}, 
\end{equation}
where $\omega_{\mathrm{c}}$ is cutoff frequency and $\eta$ controls the coupling strength. Figure \ref{Ohmic1}(\textbf{\textsf{a}}) depicts the dynamics of the excited-state population and (\textbf{\textsf{b}}) the norm of the correlation obtained by exact numerical simulations (from ULL), ULL$2$, TCL$2$, NZ$2$, Redfield, and CR approximations. It should be noted that since $H_{\mathsf{S}}$ has only one energy gap, the Lindblad and the Redfield equations are equivalent here. The initial decay for $\tau \leqslant 0.1$ of the excited-state population is accurately captured by the ULL2, NZ2 and TCL2 methods. Interestingly, the initial dynamics of the norm shows an increasing trend, indicating the growing strength of the correlation, which leads to the initial decay dynamics. This demonstrates the role of correlation in short-time dynamics. Furthermore, it is observed that the Redfield equations provide an erroneous estimate of the initial correlation [see the inset in Fig. \ref{Ohmic1}(\textbf{\textsf{b}})], resulting in inaccurate short-time dynamics. However, the corrected correlation and the resulting Redfield equation accurately replicate the short-time dynamics. In the later part of the dynamics (for $\tau>0.2)$, the norm of the exact correlation is almost constant, and the population dynamics exhibits a plateau. Although all approximative results show deviations in this time scale, ULL$2$ remains more accurate than the other techniques.

In Figs. \ref{Ohmic1}(\textbf{\textsf{c}}) and (\textbf{\textsf{d}}), we show the accumulated error in the dynamics of the state of the system and the system-environment correlation obtained by the ULL$2$, TCL$2$, NZ$2$, Redfield, and CR approximations. It is observed from the plots that the ULL$2$ equation captures the dynamics of the system more accurately than the other techniques---for the given parameter set. In addition, the ULL$2$ technique outperforms the other techniques in capturing the system-environment correlation, as can be justified by comparing the distance of the approximated correlation operators from the exact one, $D_{\textsc{hs}}(\chi^{\scriptscriptstyle{a}},\chi^{\textsc{ex}})$. Note that this distance takes into account features of the correlation operator and is a more faithful measure than the norm, which only assigns a single number to the operator, in demonstrating which technique works relatively better.

%%%%%%%%%%%%%%%%%%%%%%%%%%%%%%%%%%%%%%%%%%%%%%%%%%%%%%%%%%%%%%%%
\subsection{Lorentzian environment}
\label{sec:lorentzian}

Next we consider the Lorentzian spectral density 
\begin{equation}
J(\omega) = \frac{\Gamma \lambda^{2}}{2\pi ({\omega^{2} + \lambda^{2} })}, 
\end{equation}
where $\Gamma$ is the coupling strength and $1/\lambda$ determines the environment correlation time \cite{Maniscalco}. It has been discussed in Ref. \cite{spinqubit_wu} that when $\lambda/\Gamma<1$ one can expect some non-Markovian characteristics at a reasonably small timescale. In Figs. \ref{LO1}, we analyze the different equations for this spectral density by studying the same quantities as in Figs. \ref{Ohmic1}. In this case, NZ$2$ and CR give unphysical negative populations, and hence we did not plot them in Figs. \ref{LO1}.\\

The exact population dynamics displays an oscillatory behavior and the corresponding norm of the correlation operator exhibits a revival of correlation. This revival may be attributed to the non-Markovian nature of the bath. Similar to the previous case, we observe that the excited-state population dynamic is accurate in the ULL$2$ and TCL$2$ results in short times for $\tau <0.5$ [see Fig. \ref{LO1}(\textbf{\textsf{a}})]. The dynamics of the ULL$2$ and TCL$2$ correlations closely follow the norm of the exact correlation at short times, as depicted in Fig. \ref{LO1}(\textbf{\textsf{b}}). However, for later times $\tau >0.5$, the norm of the correlation and the excited-state population dynamics calculated using ULL$2$ and TCL$2$ start to deviate, which is expected due to the approximate nature of the associated correlation operators. Nonetheless, ULL$2$ partially captures the oscillatory dynamics of the excited-state population and the revival of the correlation. It is also seen from the inset in Fig. \ref{LO1}(\textbf{\textsf{b}}) that the Redfield equation has a nonzero initial correlation and significantly deviates from the exact correlation. Furthermore, Figs. \ref{LO1}(\textbf{\textsf{c}}) and (\textbf{\textsf{d}}) illustrate the accumulated errors in the dynamics and the correlation. These results demonstrate that ULL$2$ is capable of capturing the system-environment correlation better.

%%%%%%%%%%%%%%%%%%%%%%%%%%%%%%%%%%%%%%%%%%%%%%%%%%%%%%%%%%%%%%%%
\section{Summary and conclusions}
\label{sec:summary}

We have obtained the system-environment correlation associated with the approximate description of reduced dynamics of open quantum systems by writing reduced dynamics in a universal form wherein the system-environment correlation operator is present. This has enabled us to read off the approximate system-environment correlation from the reduced dynamics of the system and the environment within a given system-environment model. This way, we have demonstrate how to retain system-environment correlation within commonly used master equations, including the Markovian Lindblad, Redfield, second-order time-convolutionless, second-order Nakajima-Zwanzig, and second-order universal Lindblad-like equations. We have shown that each master equation is indeed associated with a particular approximation of the system-environment correlation operator. This correlation-based analysis has offered an explanation for the inconsistency of the Redfield equation at initial or short times and a remedy for that. This can be a complement to the initial-condition slippage rectification of this master equation. Moreover, we have noted that in the Lindblad equation system-environment correlations come into a system-gap decomposed structure, which has been induced by the rotating-wave approximation. We have illustrated our findings in two examples by demonstrating that ULL$2$ predicts the dynamics of the system-bath correlation and dynamics of the system and the bath more accurately than other approximate standard master equations. 

%%%%%%%%%%%%%%%%%%%%%%%%%%%%%%%%%%%%%%%%%%%%%%%%%%%%%%%%%%%%%%%%
\textit{Acknowledgments.}---This work was supported in part by the Academy of Finland QTF Center of Excellence program (Project No. 312298) and Sharif University of Technology's Office of Vice President for Research and Technology.
%%%%%%%%%%%%%%%%%%%%%%%%%%%%%%%%%%%%%%%%%%%%%%%%%%%%%%%%%%%%%%%%

%%%%%%%%%%%%%%%%%%%%%%%%%%%%%%%%%%%%%%%%%%%%%%%%%%%%%%%%%%%%%%%%

%%%%%%%%%%%%%%%%%%%%%%%%%%%%%%%%%%%%%%%%%%%%%%%%%%%%%%%%%%%%%%%%
\onecolumngrid
%%%%%%%%%%%%%%%%%%%%%%%%%%%%%%%%%%%%%%%%%%%%%%%%%%%%%%%%%%%%%%%%
\newpage
\appendix
\widetext

%%%%%%%%%%%%%%%%%%%%%%%%%%%%%%%%%%%%%%%%%%%%%%%%%%%%%%%%%%%%%%%%
\section{Derivation of the exact dynamics}
\label{app:exactD}

To see how correlation appears in the dynamical master equation of the system, we start from the Schr\"{o}dinger equation of the total system $\dot{\varrho}_{\mathsf{SB}}(\tau)=-i [H_{\mathsf{SB}},\varrho_{\mathsf{SB}}(\tau)]$. By inserting $\varrho_{\mathsf{SB}}(\tau)=\varrho_{\mathsf{S}}(\tau)\otimes \varrho_{\mathsf{B}}(\tau)+\chi(\tau)$ into the Schr\"{o}dinger equation, we obtain  
\begin{align}
\dot{\varrho}_{\mathsf{SB}}(\tau)&=-i [H_{\mathsf{SB}},\varrho_{\mathsf{S}}(\tau)\otimes \varrho_{\mathsf{B}}(\tau)]-i [H_{\mathsf{SB}}, \chi(\tau)].
\label{exactSB}
\end{align}
The exact dynamical equation of the system is obtained by tracing over the environmental degrees of freedom as
\begin{align}
\dot{\varrho}_{\mathsf{S}}(\tau)&=-i [\widetilde{H}_{\mathsf{S}}(\tau),\varrho_{\mathsf{S}}(\tau)]-i \,\mathrm{Tr}_{\mathsf{B}}[H_{\mathrm{I}},
\chi(\tau)],\label{exactS}
\end{align}
where $\widetilde{H}_{\mathsf{S}}(\tau)=H_{\mathsf{S}}+\mathrm{Tr}_{\mathsf{B}}[H_{\mathrm{I}}\varrho_{\mathsf{B}}(\tau)]$. In derivation of the above equation we have used the identity $\mathrm{Tr}_{\mathsf{B}}[H_{\mathrm{I}},\varrho_{\mathsf{S}}(\tau)\otimes \varrho_{\mathsf{B}}(\tau)]=\big[\mathrm{Tr}_{\mathsf{B}}[H_{\mathrm{I}} \varrho_{\mathsf{B}}(\tau)],\varrho_{\mathsf{S}}(\tau)\big]$, $\mathrm{Tr}_\mathsf{B} [H_{\mathsf{B}} ,\chi]=0$, and $\mathrm{Tr}_\mathsf{B}[H_{\mathsf{S}}, \chi]=[H_{\mathsf{S}},\mathrm{Tr}_\mathsf{B}[\chi]]=0$. Since working in the interaction picture is more convenient and the interaction-picture version of the approximate equations are more common, we also rewrite Eq. \eqref{exactS} in the interaction picture which leads to 
\begin{align}
 \dot{\boldsymbol{\varrho}}_{\mathsf{S}}(\tau)&=-i \big[\mathrm{Tr}_{\mathsf{B}}[\boldsymbol{H}_{\mathrm{I}}(\tau)\boldsymbol{\varrho}_{\mathsf{B}}(\tau)],\boldsymbol{\varrho}_{\mathsf{S}}(\tau)\big]-i \mathrm{Tr}_{\mathsf{B}}[\boldsymbol{H}_{\mathrm{I}}(\tau),\boldsymbol{\chi}(\tau)].
\label{exactS-intpic-}
 \end{align}
This is a universal form of the dynamical equation where the correlation operator $\boldsymbol{\chi}$ is present. This is Eq. \eqref{exactS-intpic} of the main text.

%%%%%%%%%%%%%%%%%%%%%%%%%%%%%%%%%%%%%%%%%%%%%%%%%%%%%%%%%%%%%%%%

\section{Derivation of the approximated correlation from the exact correlation}
\label{app:new}

Here we derive the approximated correlation and corresponding master equations from the exact dynamics. Consider the exact dynamical equation in the interaction picture
\begin{equation}
\dot{\boldsymbol{\varrho}}_{\mathsf{SB}}(\tau)=-i [ \boldsymbol{ H}_{\mathrm{I}}(\tau),\boldsymbol{\varrho}_{\mathsf{SB}}(\tau)].
\label{VN_int}
\end{equation}  
 The formal solution to this equation can be written as
\begin{equation}
\boldsymbol{\varrho}_{\mathsf{SB}}(\tau)=\boldsymbol{\varrho}_{\mathsf{SB}}(0)-i \textstyle{\int_{0}^{\tau}} ds\, [\boldsymbol{H}_{\mathrm{I}}(s),\boldsymbol{\varrho}_{\mathsf{SB}}(s)].
\label{formal_rhoSB}
\end{equation}
Similarly, one can also calculate the formal solution for the environment and system density operators by tracing over the system and environment degrees of freedom, respectively: 
\begin{align}
\boldsymbol{\varrho}_{\mathsf{S}}(\tau)=&\boldsymbol{\varrho}_{\mathsf{S}}(0)-i \textstyle{\int_{0}^{\tau}} ds\,\mathrm{Tr}_{\mathsf{B}} [\boldsymbol{H}_{\mathrm{I}}(s),\boldsymbol{\varrho}_{\mathsf{SB}}(s)],
\label{formal_rhoS}
\\
\boldsymbol{\varrho}_{\mathsf{B}}(\tau)=&\boldsymbol{\varrho}_{\mathsf{B}}(0)-i \textstyle{\int_{0}^{\tau}} ds\, \mathrm{Tr}_{\mathsf{S}} [\boldsymbol{H}_{\mathrm{I}}(s),\boldsymbol{\varrho}_{\mathsf{SB}}(s)].
\label{formal_rhoB}
\end{align}
Substituting these formal solutions into the interaction-picture form of Eq. \eqref{correlation} in the main text gives the exact equation
\begin{align}
\boldsymbol{\chi}(\tau)=&\boldsymbol{\chi}(0)-i \textstyle{\int_{0}^{\tau}} ds\, \big([\boldsymbol{ H}_{\mathrm{I}}(s),\boldsymbol{\varrho}_{\mathsf{SB}}(s)]-\boldsymbol{\varrho}_{\mathsf{S}}(0)\otimes \mathrm{Tr}_{\mathsf{S}}[\boldsymbol{ H}_{\mathrm{I}}(s),\boldsymbol{\varrho}_{\mathsf{SB}}(s)]-\mathrm{Tr}_{\mathsf{B}}[\boldsymbol{ H}_{\mathrm{I}}(s),\boldsymbol{\varrho}_{\mathsf{SB}}(s)]\otimes \boldsymbol{\varrho}_{\mathsf{B}}(0)\big)\nonumber\\
&- \textstyle{\int_{0}^{\tau}} ds\, \mathrm{Tr}_{\mathsf{B}}[\boldsymbol{ H}_{\mathrm{I}}(s),\boldsymbol{\varrho}_{\mathsf{SB}}(s)]\otimes \textstyle{\int_{0}^{\tau}} ds\,\mathrm{Tr}_{\mathsf{S}}[\boldsymbol{ H}_{\mathrm{I}}(s),\boldsymbol{\varrho}_{\mathsf{SB}}(s)].
\label{apporx_chi_secondtorder_HI}
\end{align}
If we set the initial condition set $\boldsymbol{\chi}(0)=\boldsymbol{\varrho}_{\mathsf{SB}}(0)-\boldsymbol{\varrho}_{\mathsf{S}}(0)\otimes \boldsymbol{\varrho}_{\mathsf{B}}(0)=0$ and consider a weak-coupling limit by neglecting the last term which is of second order in $\| \boldsymbol{H}_{\mathrm{I}} \|$, we obtain the approximated correlation as 
\begin{align}
\boldsymbol{\chi}(\tau)&\approx -i \textstyle{\int_{0}^{\tau}} ds\, \big([\boldsymbol{ H}_{\mathrm{I}}(s),\boldsymbol{\varrho}_{\mathsf{SB}}(s)]-\boldsymbol{\varrho}_{\mathsf{S}}(0)\otimes \mathrm{Tr}_{\mathsf{S}}[\boldsymbol{ H}_{\mathrm{I}}(s),\boldsymbol{\varrho}_{\mathsf{SB}}(s)]-\mathrm{Tr}_{\mathsf{B}}[\boldsymbol{ H}_{\mathrm{I}}(s),\boldsymbol{\varrho}_{\mathsf{SB}}(s)]\otimes \boldsymbol{\varrho}_{\mathsf{B}}(0)\big).
\label{apporx_chi_firstorder_HI}
\end{align}
Now we show that this correlation operator can be taken as an alternative basis for derivation of some of the standard master equations.

\begin{itemize}

\item If we apply the Born approximation $\boldsymbol{\varrho}_{\mathsf{SB}}(s)\approx \boldsymbol{\varrho}_{\mathsf{S}}(s)\otimes \boldsymbol{\varrho}_{\mathsf{B}}(0)$ inside the integral, we obtain
\begin{align}
\boldsymbol{\chi}^{\textsc{nz}\scriptscriptstyle{2}}(\tau)&\approx -i \textstyle{\int_{0}^{\tau}} ds\,  \big([\boldsymbol{H}_{\mathrm{I}}(s),\boldsymbol{\varrho}_{\mathsf{S}}(s)\otimes \boldsymbol{\varrho}_{\mathsf{B}}(0)]-\boldsymbol{\varrho}_{\mathsf{S}}(0)\otimes \mathrm{Tr}_{\mathsf{S}}[\boldsymbol{ H}_{\mathrm{I}}(s),\boldsymbol{\varrho}_{\mathsf{S}}(s)\otimes \boldsymbol{\varrho}_{\mathsf{B}}(0)]\big),
\label{apporx_born}
\end{align}
where we have used $\mathrm{Tr}_{\mathsf{B}}[\boldsymbol{H}_{\mathrm{I}}(s), \boldsymbol{\varrho}_{\mathsf{B}}(0)]=\mathrm{Tr}_{\mathsf{B}}[H_{\mathrm{I}}, \varrho_{\mathsf{B}}(0)]=0$. Substituting $\boldsymbol{\chi}^{\textsc{nz}\scriptscriptstyle{2}}(\tau)$ in Eq. \eqref{exactS-intpic} of the main text yields the NZ$2$ equation. 

\item In addition, if we also apply the Markov approximation by replacing $\boldsymbol{\varrho}_{\mathsf{S}}(s)$ with $\boldsymbol{\varrho}_{\mathsf{S}}(\tau)$, we arrive at
\begin{align}
\boldsymbol{\chi}^{\textsc{tcl}\scriptscriptstyle{2}}(\tau)& \approx-i \textstyle{\int_{0}^{\tau}} ds\, \big([\boldsymbol{ H}_{\mathrm{I}}(s),\boldsymbol{\varrho}_{\mathsf{S}}(\tau)\otimes \boldsymbol{\varrho}_{\mathsf{B}}(0)]-\boldsymbol{\varrho}_{\mathsf{S}}(0)\otimes \mathrm{Tr}_{\mathsf{S}}[\boldsymbol{ H}_{\mathrm{I}}(s),\boldsymbol{\varrho}_{\mathsf{S}}(\tau)\otimes \boldsymbol{\varrho}_{\mathsf{B}}(0)]\big).
\label{apporx_born_Markov}
\end{align}
We can recover the TCL$2$ equation by substituting $\boldsymbol{\chi}^{\textsc{tcl}\scriptscriptstyle{2}}(\tau)$ in Eq. \eqref{exactS-intpic}.
 
\item Similarly, we can obtain the approximated correlation corresponding to the Redfield equation by setting the upper limit of the integral in Eq. \eqref{apporx_chi_firstorder_HI} to be infinity and changing $s$ to $\tau-s$,
\begin{align}
\boldsymbol{\chi}^{\textsc{r}}(\tau)& \approx -i \textstyle{\int_{0}^{\infty}} ds\,  ([\boldsymbol{ H}_{\mathrm{I}}(\tau-s),\boldsymbol{\varrho}_{\mathsf{S}}(\tau)\otimes \boldsymbol{\varrho}_{\mathsf{B}}(0)]-\boldsymbol{\varrho}_{\mathsf{S}}(0)\otimes \mathrm{Tr}_{\mathsf{S}}[\boldsymbol{ H}_{\mathrm{I}}(\tau-s),\boldsymbol{\varrho}_{\mathsf{S}}(\tau)\otimes \boldsymbol{\varrho}_{\mathsf{B}}(0)]).
\label{apporx_born_Markov-}
\end{align}

\end{itemize}

As a final remark, we note that the neglected term in Eq. \eqref{apporx_chi_firstorder_HI} is not the only term of second order in $\boldsymbol{H}_{\mathrm{I}}$ in Eq. \eqref{apporx_chi_secondtorder_HI}. For a rigorous analysis of $\boldsymbol{\chi}$ order by order is provided through the ULL technique \cite{PRX}. This is, however, beyond the scope of this paper and the current approximation (\ref{apporx_chi_firstorder_HI}) suffices for the purpose of reading approximate expressions for $\boldsymbol{\chi}$.

%%%%%%%%%%%%%%%%%%%%%%%%%%%%%%%%%%%%%%%%%%%%%%%%%%%%%%%%%%%%%%%%
\section{Details of the example}
\label{app:example-details}

We consider a qubit interacting with an environment of qubits through the following Hamiltonian (in the interaction picture):  
\begin{align}
\boldsymbol{H}_{\mathrm{I}}(\tau)=\textstyle{\sum_{k=1}^{M}}g_{k} \big(\sigma_{-}\Sigma^{k}_{+}  e^{-{i}(\omega_{0}-\omega_{k})\tau}+ \sigma_{+} \Sigma^{k}_{-} e^{i(\omega_{0}-\omega_{k})\tau}\big),
\end{align}
where we have used the following relations for the operators in the interaction picture:
\begin{align}
\boldsymbol{\sigma}_{\pm}(\tau) &=\sigma_{\pm}e^{\pm i\omega_{0}\tau},\\
\boldsymbol{\Sigma}_{\pm}^{k}(\tau) &=\Sigma_{\pm}^{k} e^{\pm i\omega_{k} \tau }.
\end{align}
%%%%%%%%%%%%%%%%%%%%%%%%%%%%%%%%%%%%%%%%%%%%%%%%%%%%%%%%%%%%%%%%
\subsection{Derivation of the Redfield equation}

Consider the Redfield equation in the interaction picture, 
\begin{equation}
\dot{\boldsymbol{\varrho}}_{\mathsf{S}}(\tau)=-\textstyle{\int_{0}^{\infty}}ds\, \mathrm{Tr}_{\mathsf{B}}\big[\boldsymbol{H}_{\mathrm{I}}(\tau),[\boldsymbol{H}_{\mathrm{I}}(\tau-s), \boldsymbol{\varrho}_{\mathsf{S}}(\tau) \otimes \boldsymbol{\varrho}_{\mathsf{B}}(0)] \big]\equiv \mathpzc{L} [\boldsymbol{\varrho}_{\mathsf{S}}(\tau)].
\label{RF-equation-integralform_sm}
\end{equation}
To find the commutator in the integral,  we evaluate each terms in the commutator separately. First, 
\begin{align}
&\mathrm{Tr}_{\mathsf{B}}[\boldsymbol{H}_{\mathrm{I}}(\tau) \boldsymbol{H}_{\mathrm{I}}(\tau-s) \boldsymbol{\varrho}_{\mathsf{S}}(\tau) \otimes \boldsymbol{\varrho}_{\mathsf{B}}(0)]= \nonumber \\ &  \textstyle{\sum_{k'=1}^{M}} \textstyle{\sum_{k=1}^{M}} g_{k} g_{k}'\big(  \boldsymbol{\sigma}_{+}(\tau) \boldsymbol{\sigma}_{+}(\tau-s) \boldsymbol{\varrho}_{\mathsf{S}}(\tau) \mathrm{Tr}_{\mathsf{B}}[\boldsymbol{\Sigma}^{k'}_{-}(\tau) \boldsymbol{\Sigma}^{k}_{-}(\tau-s) \boldsymbol{\varrho}_{\mathsf{B}}(0)] + \boldsymbol{\sigma}_{-}(\tau) \boldsymbol{\sigma}_{-}(\tau-s) \boldsymbol{\varrho}_{\mathsf{S}}(\tau) \mathrm{Tr}_{\mathsf{B}}[\boldsymbol{\Sigma}^{k'}_{+}(\tau) \boldsymbol{\Sigma}^{k}_{+}(\tau-s) \boldsymbol{\varrho}_{\mathsf{B}}(0)] \nonumber\\
&+ \boldsymbol{\sigma}_{+}(\tau) \boldsymbol{\sigma}_{-}(\tau-s) \boldsymbol{\varrho}_{\mathsf{S}}(\tau) \mathrm{Tr}_{\mathsf{B}}[\boldsymbol{\Sigma}^{k'}_{-}(\tau) \boldsymbol{\Sigma}^{k}_{+}(\tau-s) \boldsymbol{\varrho}_{\mathsf{B}}(0)] + \boldsymbol{\sigma}_{-}(\tau) \boldsymbol{\sigma}_{+}(\tau-s) \boldsymbol{\varrho}_{\mathsf{S}}(\tau) \mathrm{Tr}_{\mathsf{B}}[\boldsymbol{\Sigma}^{{k'}}_{+}(\tau) \boldsymbol{\Sigma}^{k}_{-}(\tau-s) \boldsymbol{\varrho}_{\mathsf{B}}(0)]\big).
\label{eq11}
\end{align}
We further simplify the above relation using $\mathrm{Tr}_{\mathsf{B}}[\boldsymbol{\Sigma}^{k'}_{-}(\tau) \boldsymbol{\Sigma}^{k}_{-}(\tau-s)\boldsymbol{\varrho}_{\mathsf{B}}(0)]=\mathrm{Tr}_{\mathsf{B}}[\boldsymbol{\Sigma}^{k'}_{+}(\tau) \boldsymbol{\Sigma}^{k}_{+}(\tau-s) \boldsymbol{\varrho}_{\mathsf{B}}(0)]=0$ and obtain 
\begin{align}
&\mathrm{Tr}_{\mathsf{B}}[ \boldsymbol{H}_{\mathrm{I}}(\tau) \boldsymbol{H}_{\mathrm{I}}(\tau-s) \boldsymbol{\varrho}_{\mathsf{S}}(\tau) \otimes \boldsymbol{\varrho}_{\mathsf{B}}(0)]=\nonumber\\&  \textstyle{\sum_{k'=1}^{M}} \textstyle{\sum_{k=1}^{M}} g_{k} g_{k}'\big( \boldsymbol{\sigma}_{+}(\tau) \boldsymbol{\sigma}_{-}(\tau-s) \boldsymbol{\varrho}_{\mathsf{S}}(\tau) \mathrm{Tr}_{\mathsf{B}}[\boldsymbol{\Sigma}^{k'}_{-}(\tau) \boldsymbol{\Sigma}^{k}_{+}(\tau-s) \boldsymbol{\varrho}_{\mathsf{B}}(0)] + \boldsymbol{\sigma}_{-}(\tau) \boldsymbol{\sigma}_{+}(\tau-s) \boldsymbol{\varrho}_{\mathsf{S}}(\tau) \mathrm{Tr}_{\mathsf{B}}[\boldsymbol{\Sigma}^{k'}_{+}(\tau) \boldsymbol{\Sigma}^{k}_{-}(\tau-s) \boldsymbol{\varrho}_{\mathsf{B}}(0)] \big) =\nonumber\\
& \textstyle{\sum_{k'=1}^{M}} \textstyle{\sum_{k=1}^{M}} g_{k} g_{k}' \big( \sigma_{+}\sigma_{-}e^{i \omega_{0}s} \boldsymbol{\varrho}_{\mathsf{S}}(\tau) \mathrm{Tr}_{\mathsf{B}}[\Sigma^{k'}_{-}\Sigma^{k}_{+} \boldsymbol{\varrho}_{\mathsf{B}}(0)]e^{-i (\omega_{k'} \tau-\omega_{k} (\tau-s))}+\sigma_{-}\sigma_{+} e^{-i \omega_{0}(s)} \boldsymbol{\varrho}_{\mathsf{S}}(\tau) \mathrm{Tr}_{\mathsf{B}}[\Sigma^{k'}_{+}\Sigma^{k}_{-}\varrho_{\mathsf{B}}(0)]e^{i (\omega_{k'} \tau-\omega_{k} (\tau-s))}\big).
\label{eq12}
\end{align}
We assume that the environment is initially in the pure state $|0\rangle_{\mathsf{B}}^{\otimes M}$, from which   
\begin{equation}
\begin{split}
\mathrm{Tr}_{\mathsf{B}}[\boldsymbol{H}_{\mathrm{I}}(\tau) \boldsymbol{H}_{\mathrm{I}}(\tau-s) \boldsymbol{\varrho}_{\mathsf{S}}(\tau) \otimes \boldsymbol{\varrho}_{\mathsf{B}}(0)] = \textstyle{\sum_{k=1}^{M}} g_{k}^{2} \sigma_{+} \sigma_{-} \boldsymbol{\varrho}_{\mathsf{S}}(\tau) e^{i(\omega_{0}-\omega_{k})s} = L(s)\, \sigma_{+} \sigma_{-} \boldsymbol{\varrho}_{\mathsf{S}}(\tau),
\end{split}
\label{eq13}
\end{equation}
where 
\begin{equation}
L(s)\equiv\textstyle{\sum_{k=1}^{M}} g_{k}^{2} e^{i(\omega_{0}-\omega_{k})s}. 
\end{equation}
Similarly one can calculate other terms in the integral as  
\begin{align}
& \mathrm{Tr}_{\mathsf{B}}[\boldsymbol{H}_{\mathrm{I}}(\tau) \boldsymbol{\varrho}_{\mathsf{S}}(\tau) \otimes \boldsymbol{\varrho}_{\mathsf{B}}(0) \boldsymbol{H}_{\mathrm{I}}(\tau-s)] = \textstyle{\sum_{k=1}^{M}} g_{k}^{2} \sigma_{-} \boldsymbol{\varrho}_{\mathsf{S}}(\tau)\sigma_{+}  e^{-i(\omega_{0}-\omega_{k})s} = L(-s)\, \sigma_{-} \boldsymbol{\varrho}_{\mathsf{S}}(\tau)\sigma_{+},\\
& \mathrm{Tr}_{\mathsf{B}}[\boldsymbol{H}_{\mathrm{I}}(\tau-s) \boldsymbol{\varrho}_{\mathsf{S}}(\tau) \otimes \boldsymbol{\varrho}_{\mathsf{B}}(0) \boldsymbol{H}_{\mathrm{I}}(\tau)] = \textstyle{\sum_{k=1}^{M}} g_{k}^{2} \sigma_{-} \boldsymbol{\varrho}_{\mathsf{S}}(\tau)\sigma_{+}  e^{i(\omega_{0}-\omega_{k})s} = L(s)\, \sigma_{-} \boldsymbol{\varrho}_{\mathsf{S}}(\tau)\sigma_{+},\\
&  \mathrm{Tr}_{\mathsf{B}}[ \boldsymbol{\varrho}_{\mathsf{S}}(\tau) \otimes \boldsymbol{\varrho}_{\mathsf{B}}(0) \boldsymbol{H}_{\mathrm{I}}(\tau-s) \boldsymbol{H}_{\mathrm{I}}(\tau)] = \textstyle{\sum_{k=1}^{M}} g_{k}^{2} \boldsymbol{\varrho}_{\mathsf{S}}(\tau)\sigma_{+} \sigma_{-}   e^{-i(\omega_{0}-\omega_{k})(\tau-s)} = L(-s)\, \boldsymbol{\varrho}_{\mathsf{S}}(\tau)\sigma_{+} \sigma_{-}. 
\label{eq14}
\end{align}
Thus,
\begin{align}
\dot{\boldsymbol{\varrho}}_{\mathsf{S}}(\tau)=& -\textstyle{\int_{0}^{\infty}}  ds \,\big( L(s)\sigma_{+} \sigma_{-} \boldsymbol{\varrho}_{\mathsf{S}}(\tau) -L(-s)\sigma_{-} \boldsymbol{\varrho}_{\mathsf{S}}(\tau)\sigma_{+}-L(s) \sigma_{-} \boldsymbol{\varrho}_{\mathsf{S}}(\tau)\sigma_{+} + L(-s) \boldsymbol{\varrho}_{\mathsf{S}}(\tau)\sigma_{+} \sigma_{-}\big).
\label{eq05}
\end{align}

We write $\textstyle{\int_{0}^{\infty}}ds\, L(\pm s) =\gamma^{{\textsc{r}}} \pm i\epsilon^{{\textsc{r}}}$, where $\gamma^{{\textsc{r}}} = \mathrm{Re}\big(\textstyle{\int_{0}^{\infty}} ds\, L(s)\big)$ and $\epsilon^{{\textsc{r}}}=\mathrm{Im}\big(\textstyle{\int_{0}^{\infty}} ds\, L(s)\big)$. Then we can write the master equation as  
\begin{equation}
\begin{split}
\dot{\boldsymbol{\varrho}}_{\mathsf{S}}(\tau)= -i\epsilon^{{\textsc{r}}}[ \sigma_{+} \sigma_{-},\boldsymbol{\varrho}_{\mathsf{S}}(\tau) ] + \gamma^{{\textsc{r}}} [2\sigma_{-} \boldsymbol{\varrho}_{\mathsf{S}}(\tau)\sigma_{+}-\{\sigma_{+} \sigma_{-}, \boldsymbol{\varrho}_{\mathsf{S}}(\tau)\}].
\end{split}
\label{eq15}
\end{equation}
After transforming back to the Schr\"{o}dinger picture, we obtain 
\begin{equation}
\begin{split}
\dot{\varrho}_{\mathsf{S}}(\tau)=-i(\omega_{0}+\epsilon^{{\textsc{r}}})[\sigma_{+} \sigma_{-},\varrho_{\mathsf{S}}(\tau)] +\gamma^{{\textsc{r}}} [2\sigma_{-} \varrho_{\mathsf{S}}(\tau)\sigma_{+}-\{\sigma_{+} \sigma_{-}, \varrho_{\mathsf{S}}(\tau)\}]=\mathpzc{L} [\varrho_{\mathsf{S}}(\tau)].
\end{split}
\label{eq06}
\end{equation}
To find the exact expression for the rates, we first consider
\begin{align}
\textstyle{\int_{0}^{\infty}}ds\, L(\pm s) =  \textstyle{\int_{0}^{\infty}} ds\, \textstyle{\sum_{k=1}^{M}} g_{k}^{2} e^{i(\omega_{0}-\omega_{k})s}.
\end{align} 
Using the spectral density 
\begin{equation}
J(\omega)= \textstyle{\sum_{k}} g_{k}^{2} \delta(\omega-\omega_{k})
\end{equation}
and the integral identity 
\begin{equation}
\textstyle{\int_{0}^{\infty}}dx\, e^{ixy} = \pi \delta(y)+ i/y,
\label{dirac}
\end{equation}
we obtain 
\begin{align}
\textstyle{\int_{0}^{\infty}}ds\, L(\pm s) =  \textstyle{\int_{0}^{\infty}}ds  \textstyle{\int_{0}^{\infty}} d\omega\,  J(\omega)\, e^{i(\omega_{0}-\omega)s} =  \pi \textstyle{\int_{0}^{\infty}}d\omega\,  J(\omega)\,\delta(\omega_{0}-\omega) +{i}\textstyle{\int_{0}^{\infty}}d\omega\,  J(\omega)/(\omega_{0}-\omega).
\end{align} 
From this we can read the rates as 
\begin{gather}
\gamma^{{\textsc{r}}} =\pi J(\omega_{0}),\\
\epsilon^{{\textsc{r}}} =\textstyle{\int_{0}^{\infty}}d\omega\,  J(\omega)/(\omega_{0}-\omega).
\end{gather}

%%%%%%%%%%%%%%%%%%%%%%%%%%%%%%%%%%%%%%%%%%%%%%%%%%%%%%%%%%%%%%%%
\subsection{Derivation of the system-environment correlation in the Redfield equation}

Consider the correlation operator
\begin{align}
\boldsymbol{\chi}^{\textsc{r}}(\tau)=&-i\textstyle{\int_{0}^{\infty}}  ds \, \big([\boldsymbol{H}_{\mathrm{I}}(\tau-s),\boldsymbol{\varrho}_{\mathsf{S}}(\tau)\otimes \boldsymbol{\varrho}_{\mathsf{B}}(0)] -  \boldsymbol{\varrho}_{\mathsf{S}}(0)  \otimes\mathrm{Tr}_{\mathsf{S}}[\boldsymbol{H}_{\mathrm{I}}(\tau-s),\boldsymbol{\varrho}_{\mathsf{S}}(\tau)\otimes \boldsymbol{\varrho}_{\mathsf{B}}(0)]\big). 
\label{corr-RF-a}
\end{align}

First, we evaluate the integral $K(\tau) \equiv \textstyle{\int_{0}^{\infty}}  ds\, \boldsymbol{H}_{\mathrm{I}}(\tau-s)$. Substituting the interaction Hamiltonian gives  
\begin{align}
K(\tau)=\textstyle{\sum_{k=1}^{M}}g_{k} \big(\sigma_{-}\Sigma^{k}_{+} \,\textstyle{\int_{0}^{\infty}}  ds\, e^{-{i}(\omega_{0}-\omega_{k})(\tau-s)} + \sigma_{+} \Sigma^{k}_{-} \,\textstyle{\int_{0}^{\infty}}  ds \,e^{i(\omega_{0}-\omega_{k})(\tau-s)} \big).
\end{align}
We further simplify the above equation using the integral relation (\ref{dirac}) and obtain 
\begin{align}
K(\tau)&=\textstyle{\sum_{k=1}^{M}}\pi g_{k} \delta (\omega_{0}-\omega_{k})[F_{k}(\tau)+ F^{\dag}_{k}(\tau)]+i \textstyle{\sum_{k=1}^{M}}  \frac{ g_{k}}{(\omega_{0}-\omega_{k})} [F_{k}(\tau)-F^{\dag}_{k}(\tau)],
\end{align}
where $F_{k} \equiv \sigma_{-} \otimes \big(\mathbbmss{I}^{(1)}\otimes \mathbbmss{I}^{(2)} \otimes \cdots\otimes \Sigma^{k}_{+} \otimes \cdots \otimes \mathbbmss{I}^{(M)}\big)$. To evaluate the correlation operator numerically, we further assume an approximate form for the Dirac delta function $\delta(x) \approx e^{-x^{2} /b^{2} }/(\sqrt{\pi} |b|)$, with small values of $b$. By inserting this expression for $K(\tau)$ in Eq. \eqref{corr-RF-a}, we obtain 
\begin{align}
\boldsymbol{\chi}^{\textsc{r}}(\tau) = [K(\tau),\boldsymbol{\varrho}_{\mathsf{S}}(\tau)\otimes \boldsymbol{\varrho}_{\mathsf{B}}(0)]
-  \boldsymbol{\varrho}_{\mathsf{S}}(0) \otimes\mathrm{Tr}_{\mathsf{S}}[K(\tau),\boldsymbol{\varrho}_{\mathsf{S}}(\tau)\otimes \boldsymbol{\varrho}_{\mathsf{B}}(0)].
\label{corr-K_RF}
\end{align}

%%%%%%%%%%%%%%%%%%%%%%%%%%%%%%%%%%%%%%%%%%%%%%%%%%%%%%%%%%%%%%%%
\subsection{Derivation of the corrected Redfield equation}

Consider the Redfield equation modified with the corrected correlation operator as 
\begin{align}
\dot{\boldsymbol{\varrho}}_{\mathsf{S}}(\tau)&=-\textstyle{\int_{0}^{\infty}}ds\,\big( \mathrm{Tr}_{\mathsf{B}}[\boldsymbol{H}_{\mathrm{I}}(\tau),[ \boldsymbol{H}_{\mathrm{I}}(\tau-s), \boldsymbol{\varrho}_{\mathsf{S}}(\tau) \otimes \boldsymbol{\varrho}_{\mathsf{B}}(0)]- \mathrm{Tr}_{\mathsf{B}}[\boldsymbol{H}_{\mathrm{I}}(\tau),[\boldsymbol{H}_{\mathrm{I}}(-s),\boldsymbol{\varrho}_{\mathsf{S}}(0) \otimes \boldsymbol{\varrho}_{\mathsf{B}}(0)]\big) \nonumber\\
&\equiv \mathpzc{L}[\boldsymbol{\varrho}_{\mathsf{S}}(\tau)] - \mathpzc{L}^{\mathrm{c}} [\boldsymbol{\varrho}_{\mathsf{S}}(0)].
\label{RFC-equation-integralform_sm}
\end{align}

%%%%%%%%%%%%%%%%%%%%%%%%%%%%%%%%%%%%%%%%%%%%%%%%%%%%%%%%%%%%%%%%
\begin{figure*}[tp]
\includegraphics[width=6cm]{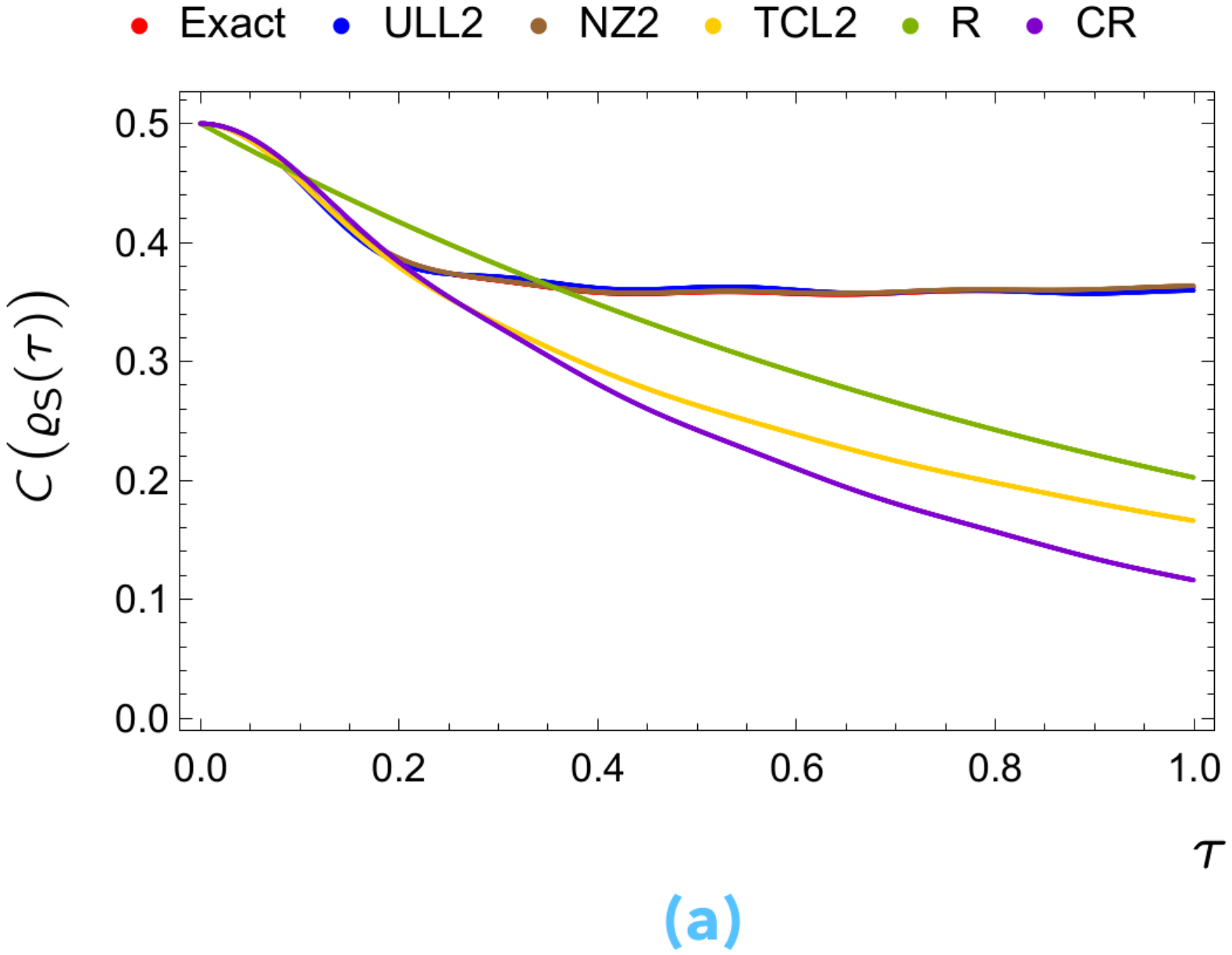}
\hskip10mm
\includegraphics[width=6cm]{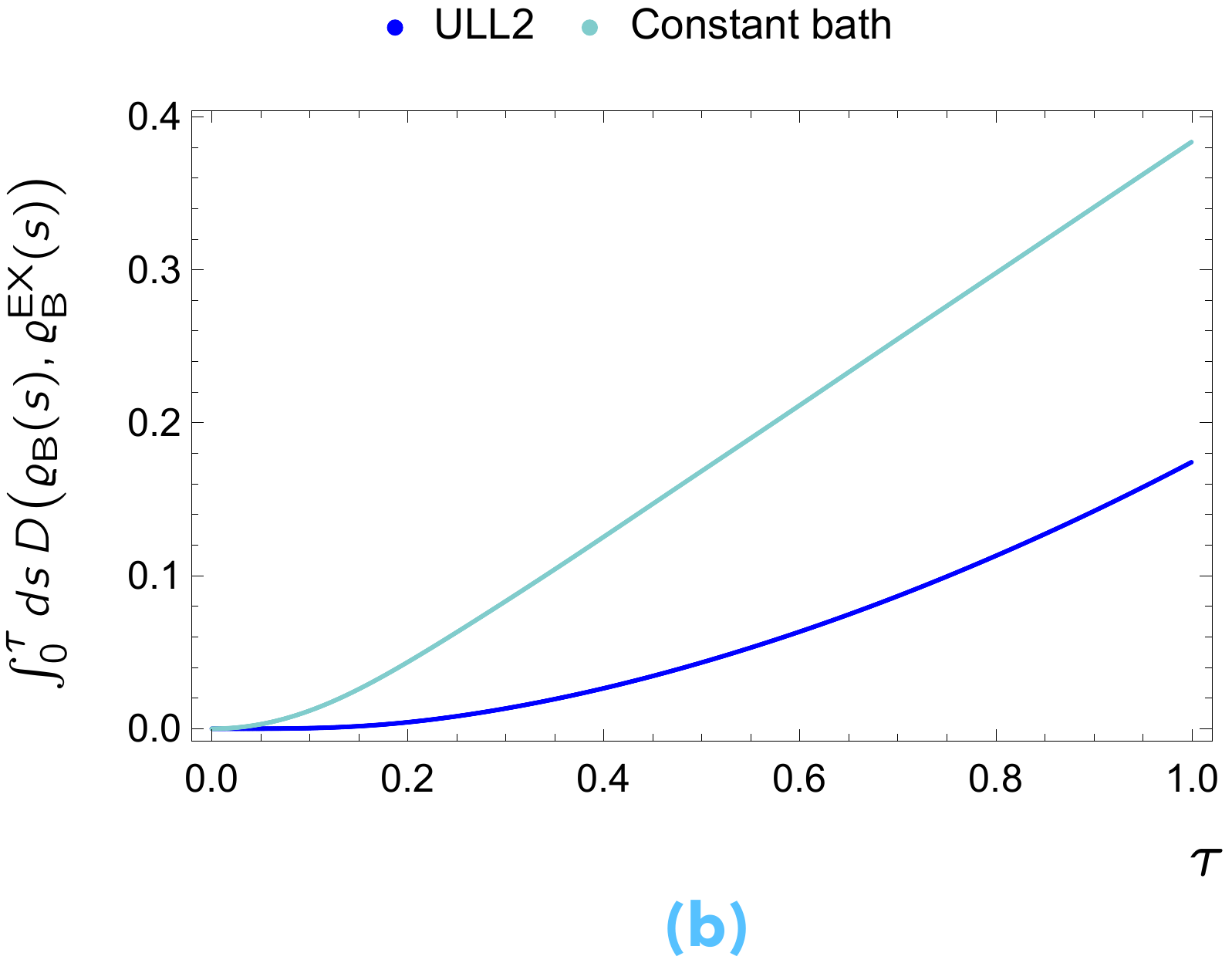}
\caption{Qubit in an Ohmic environment of qubits. (\textbf{\textsf{a}}) Dynamics of the coherence $C(\varrho_{\mathsf{S}})$ \textit{vs.} time in the exact (here ``Exact'' \& ``EX'' $\equiv$ ``ULL''), ULL$2$, NZ$2$, TCL$2$, Redfield (``R''), and corrected Redfield (``CR'') master equations. (\textbf{\textsf{b}}) Accumulative error in the approximated state of the environment. In all techniques except ULL$2$ a constant state is assumed for the environment as $\varrho_{\mathsf{B}}(\tau) = \varrho_{\mathsf{B}}(0)$. In these plots $M=255$, $\omega_{\mathrm{c}}/\omega_{0}=10$, $\eta/\omega_{0}=1$, $\Delta \omega=0.1$, and $\delta \tau=0.0005$, when the initial state of the system is $(|0\rangle+|1\rangle)/\sqrt{2}$. All quantities are in natural units where $\hbar\equiv k_{B} \equiv 1$.}
\label{correlationGamma1initialstate-SM}
\end{figure*}
%%%%%%%%%%%%%%%%%%%%%%%%%%%%%%%%%%%%%%%%%%%%%%%%%%%%%%%%%%%%%%%%

To find the dynamical equation, we first calculate the terms in the correction part $\mathpzc{L}^{\mathrm{c}} [\boldsymbol{\varrho}_{\mathsf{S}}(0)]$,
\begin{align}
&\mathrm{Tr}_{\mathsf{B}}[\boldsymbol{H}_{\mathrm{I}}(\tau)\, \boldsymbol{H}_{\mathrm{I}}(-s)\, \boldsymbol{\varrho}_{\mathsf{S}}(0) \otimes \boldsymbol{\varrho}_{\mathsf{B}}(0)]\nonumber\\& = \textstyle{\sum_{k'=1}^{M}} \textstyle{\sum_{k=1}^{M}} g_{k} g'_{k}\big(\boldsymbol{\sigma}_{+}(\tau) \boldsymbol{\sigma}_{-}(-s) \boldsymbol{\varrho}_{\mathsf{S}}(0) \,\mathrm{Tr}_{\mathsf{B}}[ \boldsymbol{\Sigma}^{k'}_{-}(\tau) \boldsymbol{\Sigma}^{k}_{+}(-s) \boldsymbol{\varrho}_{\mathsf{B}}(0)]+ \boldsymbol{\sigma}_{-}(\tau)  \boldsymbol{\sigma}_{+}(-s) \boldsymbol{\varrho}_{\mathsf{S}}(0) \,\mathrm{Tr}_{\mathsf{B}}[ \boldsymbol{\Sigma}^{k'}_{+}(\tau) \boldsymbol{\Sigma}^{k}_{-}(-s) \boldsymbol{\varrho}_{\mathsf{B}}(0)] \big) \nonumber\\
&= \textstyle{\sum_{k'=1}^{M}} \textstyle{\sum_{k=1}^{M}} g_{k}g'_{k} \big(\sigma_{+}\sigma_{-}e^{i \omega_{0}(s+\tau)} \boldsymbol{\varrho}_{\mathsf{S}}(0) \,\mathrm{Tr}_{\mathsf{B}}[\Sigma^{k'}_{-}\Sigma^{k}_{+} \boldsymbol{\varrho}_{\mathsf{B}}(0)] e^{-i (\omega_{k'} \tau+\omega_{k} s)}+\sigma_{-}\sigma_{+} e^{-i \omega_{0}(s+\tau)} \boldsymbol{\varrho}_{\mathsf{S}}(0) \,\mathrm{Tr}_{\mathsf{B}}[\Sigma^{k'}_{+}\Sigma^{k}_{-} \boldsymbol{\varrho}_{\mathsf{B}}(0)]e^{i (\omega_{k'} \tau + \omega_{k} s)}\big).
\label{eq22}
\end{align}
We assume that the environment is initially in the pure state $|0\rangle_{\mathsf{B}}^{\otimes M}$ and obtain  
\begin{equation}
\begin{split}
\mathrm{Tr}_{\mathsf{B}}[ \boldsymbol{H}_{\mathrm{I}}(\tau) \boldsymbol{H}_{\mathrm{I}}(-s) \boldsymbol{\varrho}_{\mathsf{S}}(0) \otimes \boldsymbol{\varrho}_{\mathsf{B}}(0)] = \textstyle{\sum_{k=1}^{M}} g_{k}^{2} \sigma_{+} \sigma_{-} \boldsymbol{\varrho}_{\mathsf{S}}(0) \,e^{i(\omega_{0}-\omega_{k})(s+\tau)} = L(s+\tau)\sigma_{+} \sigma_{-} \boldsymbol{\varrho}_{\mathsf{S}}(0).
\end{split}
\label{eq23}
\end{equation}
Similarly,   
\begin{align}
& \mathrm{Tr}_{\mathsf{B}}[ \boldsymbol{H}_{\mathrm{I}}(\tau) \boldsymbol{\varrho}_{\mathsf{S}}(0) \otimes \boldsymbol{\varrho}_{\mathsf{B}}(0) \boldsymbol{H}_{\mathrm{I}}(-s)] = \textstyle{\sum_{k=1}^{M}} g_{k}^{2} \sigma_{-} \boldsymbol{\varrho}_{\mathsf{S}}(\tau) \sigma_{+}  e^{-i(\omega_{0}-\omega_{k})(s+\tau)} = L(-s-\tau )\sigma_{-} \boldsymbol{\varrho}_{\mathsf{S}}(0)\sigma_{+},\\
& \mathrm{Tr}_{\mathsf{B}}[ \boldsymbol{H}_{\mathrm{I}}(-s) \boldsymbol{\varrho}_{\mathsf{S}}(0) \otimes \boldsymbol{\varrho}_{\mathsf{B}}(0) \boldsymbol{H}_{\mathrm{I}}(\tau)] = \textstyle{\sum_{k=1}^{M}} g_{k}^{2} \sigma_{-} \boldsymbol{\varrho}_{\mathsf{S}}(0) \sigma_{+}  e^{i(\omega_{0}-\omega_{k})(s+\tau)} = L(s+\tau) \sigma_{-} \boldsymbol{\varrho}_{\mathsf{S}}(0)\sigma_{+},\\
&  \mathrm{Tr}_{\mathsf{B}}[ \boldsymbol{\varrho}_{\mathsf{S}}(0) \otimes \boldsymbol{\varrho}_{\mathsf{B}}(0) \boldsymbol{H}_{\mathrm{I}}(-s) \boldsymbol{H}_{\mathrm{I}}(\tau)] = \textstyle{\sum_{k=1}^{M}} g_{k}^{2}  \boldsymbol{\varrho}_{\mathsf{S}}(0)\sigma_{+} \sigma_{-}   e^{-i(\omega_{0}-\omega_{k})(s+\tau)} = L(-s-\tau) \boldsymbol{\varrho}_{\mathsf{S}}(0)\sigma_{+} \sigma_{-}. 
\label{eq24}
\end{align}
Putting these results together yields
\begin{align}
\mathpzc{L}^{\mathrm{c}} [\varrho_{\mathsf{S}}(0)]&= -\textstyle{\int_{0}^{\infty}}  ds \,\big( L(s+\tau)\sigma_{+} \sigma_{-} \varrho_{\mathsf{S}}(0) -L(-s-\tau)\sigma_{-} \varrho_{\mathsf{S}}(0)\sigma_{+}-L(s+\tau) \sigma_{-} \varrho_{\mathsf{S}}(0)\sigma_{+}+L(-s-\tau) \varrho_{\mathsf{S}}(0)\sigma_{+} \sigma_{-}\big)\nonumber   \\& = -i\epsilon^{\mathrm{c}}_{\mathrm{R}}(\tau)[ \sigma_{+} \sigma_{-},\varrho_{\mathsf{S}}(\tau) ] + \gamma^{\mathrm{c}}_{\mathrm{R}}(\tau) [2\sigma_{-} \varrho_{\mathsf{S}}(\tau)\sigma_{+}-\{\sigma_{+} \sigma_{-}, \varrho_{\mathsf{S}}(\tau)\}],
\label{eq25}
\end{align}
where $\gamma^{{\textsc{cr}}}(\tau)=\mathrm{Re}\big(\textstyle{\int_{0}^{\infty}}ds\, L(s+\tau) \big)$ and $\epsilon^{{\textsc{cr}}}(\tau)=\mathrm{Im}\big(\textstyle{\int_{0}^{\infty}} ds\, L(s+\tau)\big)$. To find the rates, we evaluate the integral
\begin{align}
\textstyle{\int_{0}^{\infty}}ds\, L(s+\tau) &=  \textstyle{\int_{0}^{\infty}}ds  \textstyle{\int_{0}^{\infty}}d\omega\, J(\omega) \,e^{i(\omega_{0}-\omega)(s+\tau)} \,ds=  \pi \textstyle{\int_{0}^{\infty}}d\omega\, \delta(\omega_{0}-\omega)J(\omega)e^{i(\omega_{0}-\omega) \tau} +{i}\textstyle{\int_{0}^{\infty}} d\omega\, J(\omega) e^{i(\omega_{0}-\omega)\tau}/(\omega_{0}-\omega) \nonumber\\
&=\pi J(\omega_{0})-\textstyle{\int_{0}^{\infty}} d\omega\, J(\omega) \sin[(\omega_{0}-\omega)\tau]/(\omega_{0}-\omega) + i\textstyle{\int_{0}^{\infty}}d\omega\, J(\omega) \cos[(\omega_{0}-\omega)\tau]/(\omega_{0}-\omega), \nonumber
\end{align}
hence
\begin{gather}
\gamma^{{\textsc{cr}}}(\tau)=\pi J(\omega_{0})-\textstyle{\int_{0}^{\infty}}d\omega\, J(\omega) \sin[(\omega_{0}-\omega)\tau]/(\omega_{0}-\omega),\\
\epsilon^{{\textsc{cr}}}(\tau)={i}\textstyle{\int_{0}^{\infty}}d\omega\, J(\omega) \cos[(\omega_{0}-\omega)\tau]/(\omega_{0}-\omega).
\end{gather}

%%%%%%%%%%%%%%%%%%%%%%%%%%%%%%%%%%%%%%%%%%%%%%%%%%%%%%%%%%%%%%%%
\subsection{Ohmic environment}

In Figs. \ref{correlationGamma1initialstate-SM} we show the dynamics of the coherence $C\big(\varrho_{\mathsf{S}}(\tau)\big)=|\langle 0|\varrho_{\mathsf{S}}(\tau) |1\rangle|$ \cite{coherence} (where $|0\rangle$ and $|1\rangle$ form the computational basis), and the accumulative error in the approximated state of the environment in different techniques such as the exact dynamics, ULL$2$, TCL$2$, NZ$2$, Redfield, and CR approximations. It should be noted that, in this example, since $H_{\mathsf{S}}$ consists of only one energy gap, the Lindblad equation is equivalent to the Redfield equation. It is seen from the plots that the ULL$2$ equation captures both the population and coherence, \textit{i.e.}, the dynamics of the system, more accurately than the other techniques for the given parameters set. It is also observed from Figs. \ref{Ohmic1} that the system-environment correlation has also been captured more precisely by ULL$2$, compared to the other techniques.  In Figs. \ref{correlationGamma1initialstate-SM} we also show, through calculation of the integrated trace distance for the environment, $\textstyle{\int_{0}^{\tau}} ds\, D\big(\varrho^{{\scriptscriptstyle{a}}}_{\mathsf{B}}{(s)}, \varrho_{\mathsf{B}}^{\textsc{ex}}(s)\big)$, that the state of the environment deviates from the constant state approximation used in the NZ$2$, TCL$2$, Redfield, and Lindblad techniques. The environment state also evolves to a constant state that is different from the initial state and this causes a steady increase of the integrated {error}.

%%%%%%%%%%%%%%%%%%%%%%%%%%%%%%%%%%%%%%%%%%%%%%%%%%%%%%%%%%%%%%%%
\subsection{Lorentzian environment}

\begin{figure*}[tp]
\includegraphics[width=5.6cm]{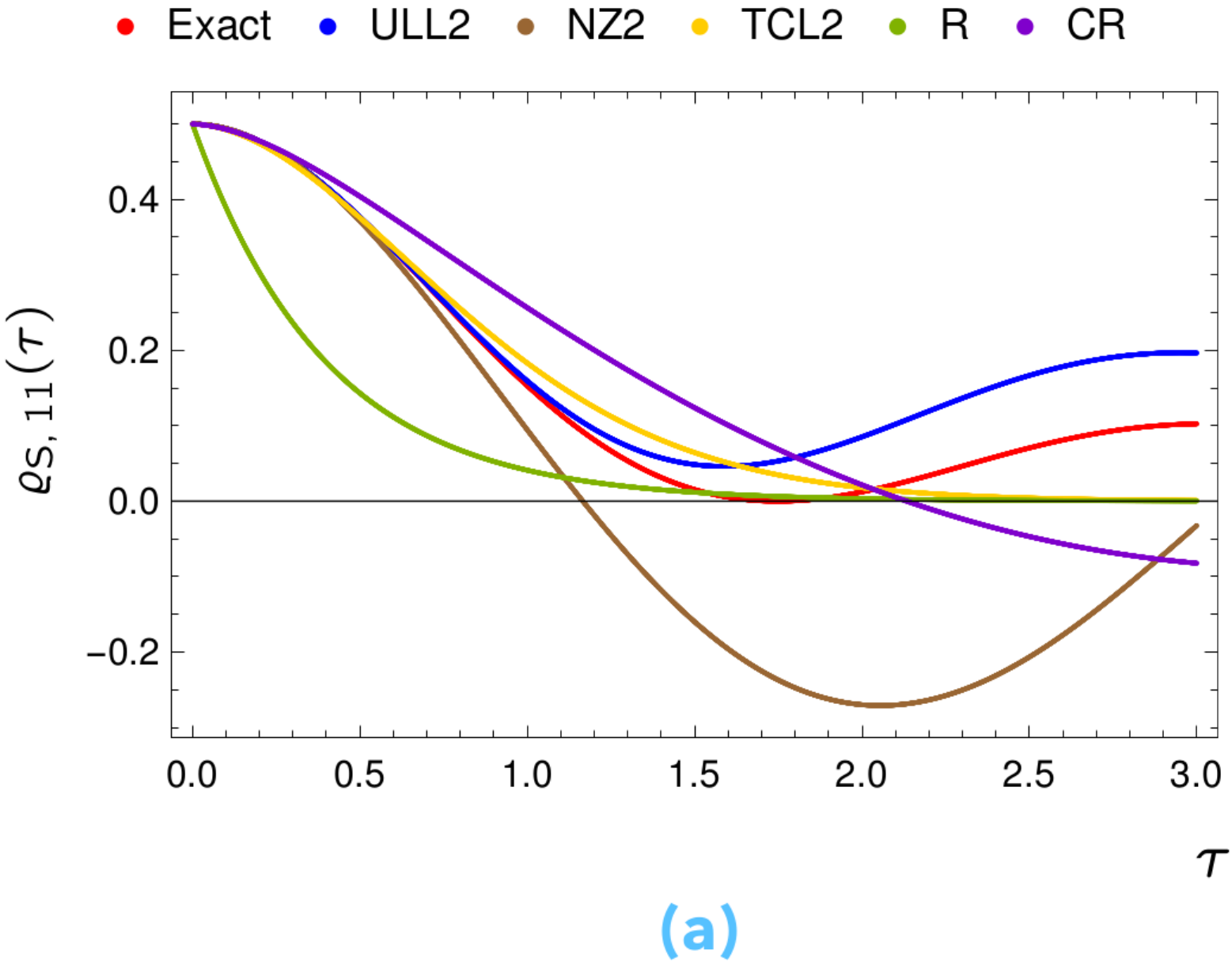} \hskip4mm 
\includegraphics[width=5.6cm]{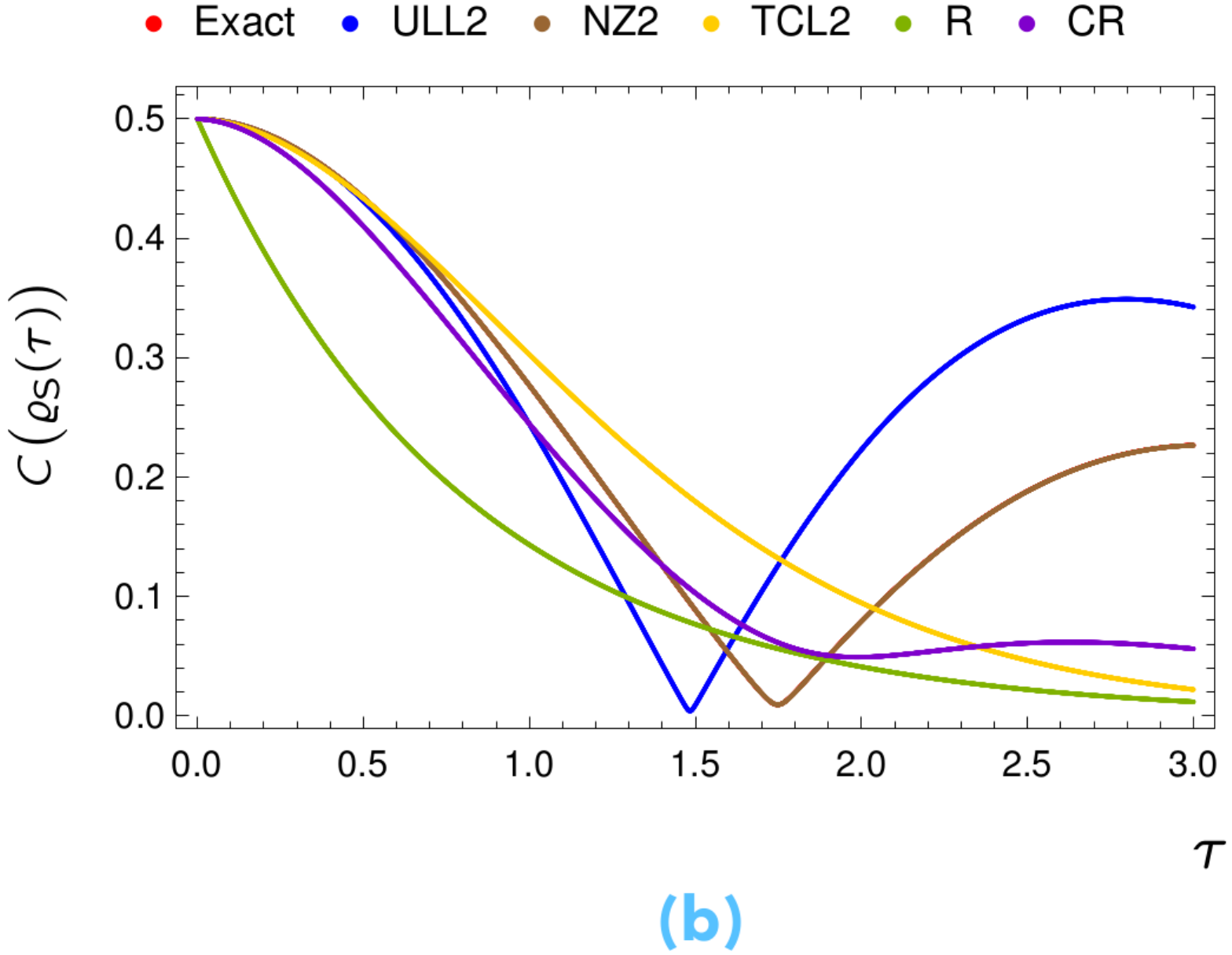} \hskip4mm 
\includegraphics[width=5.6cm]{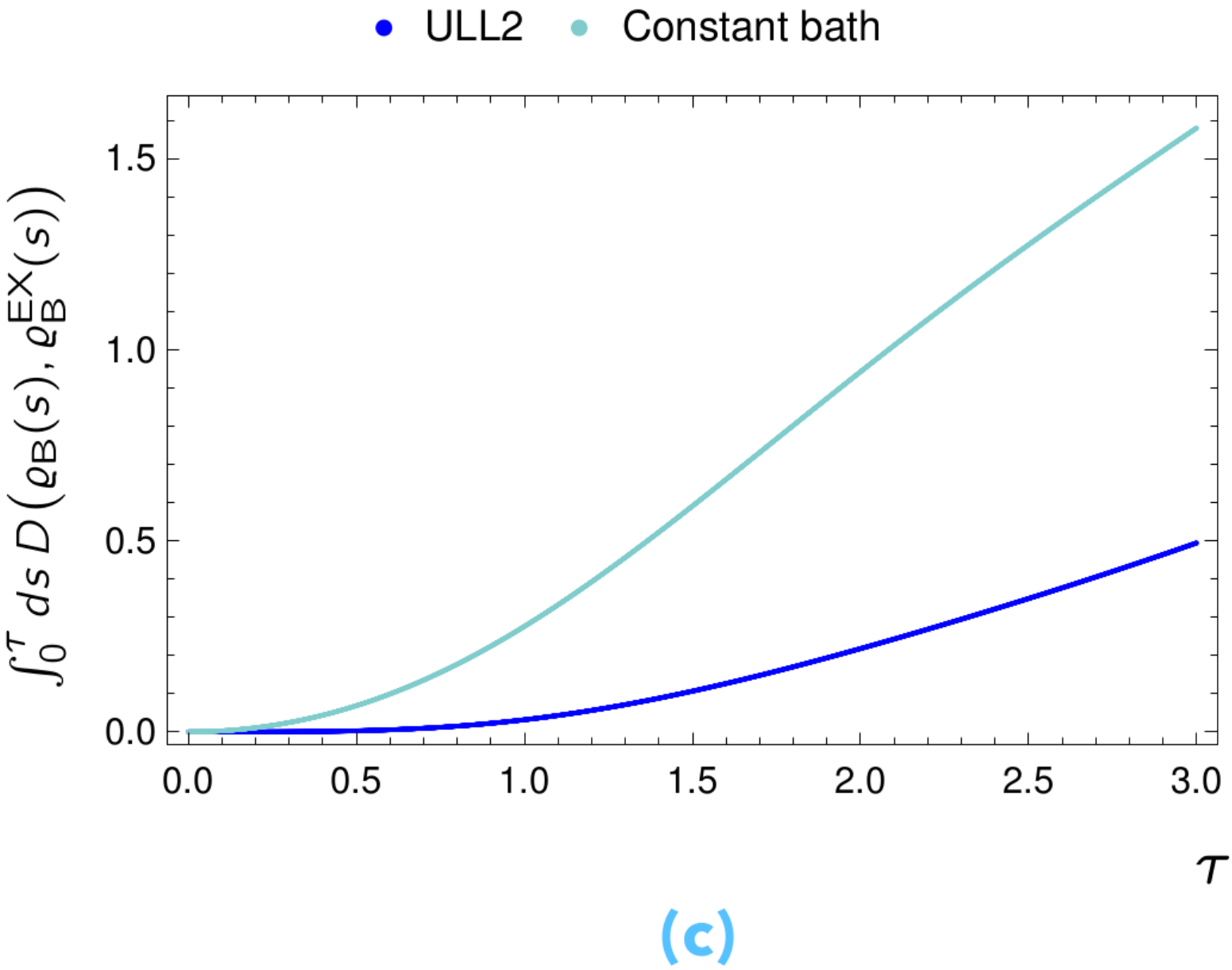}
\caption{Qubit in a Lorentzian environment of qubits. (\textbf{\textsf{a}}) Dynamics of the excited-state population $\varrho_{\mathsf{S},11}$ and (\textbf{\textsf{b}}) coherence $C(\varrho_{\mathsf{S}})$ \textit{vs.} time in the exact (here ``Exact'' \& ``EX'' $\equiv$ ``ULL''), ULL$2$, NZ$2$, TCL$2$, Redfield (``R''), and corrected Redfield (``CR'') master equations. Here the exact and NZ$2$ curves overlap, but the ULL$2$ value shows a relative phase difference. (\textbf{\textsf{c}}) Accumulative error in the approximated state of the environment. In all techniques except ULL$2$ a constant state is assumed for the environment as $\varrho_{\mathsf{B}}(\tau)=\varrho_{\mathsf{B}}(0)$. In these plots $M=255$, $\lambda/\Gamma=0.2$ , $\Delta\omega=0.05$, and $\delta \tau=0.0005$, and the initial state of the qubit is $(|0\rangle+|1\rangle)/\sqrt{2}$. All quantities are in natural units where $\hbar\equiv k_{B} \equiv 1$.}
\label{correlationlamdabyGamma0.2initialstate-SM}
\end{figure*}

As seen in Figs. \ref{correlationlamdabyGamma0.2initialstate-SM}, NZ$2$ and CR give unphysical results. The ULL$2$ technique captures the oscillatory character of the coherence, although it shows a phase difference with the exact {solution}. 

%%%%%%%%%%%%%%%%%%%%%%%%%%%%%%%%%%%%%%%%%%%%%%%%%%%%%%%%%%%%%%%%
%%%%%%%%%%%%%%%%%%%%%%%%%%%%%%%%%%%%%%%%%%%%%%%%%%%%%%%%%%%%%%%%

%%%%%%%%%%%%%%%%%%%%%%%%%%%%%%%%%%%%%%%%%%%%%%%%%%%%%%%%%%%%%%%%
\section{Example 2: Jaynes-Cummings model}
\label{app:JC}

We consider the Jaynes–Cummings model, a qubit interacting with a single cavity mode. The Hamiltonian can be written as 
\begin{equation}
H_{\mathsf{SB}}=\omega_{0} \sigma_{+}\sigma_{-}+\omega_{\mathrm{c}} a^{\dag} a+\Omega (\sigma_{+} a+\sigma_{-}a^{\dag}),
\end{equation}
where $a^{\dag}$ ($a$) are the bosonic raising (lowering) operators associated with the cavity mode and $\Omega$ determines the strength of the interaction. Note that the calculation of the correlation operator  requires a finite environment approximation. Thus, we truncate into the subspace spanned by $N$ lowest energy eigenstates of the single-mode environment. We further assume that the environment is initially  in the thermal state $\varrho_{\mathsf{B}}(0)=e^{-\beta \omega_{\mathrm{c}} a^{\dag} a}/{ \mathrm{Tr}[e^{-\beta \omega_{\mathrm{c}} a^{\dag} a}]}$, where $\beta=1/T$ with $T$ being the temperature of the environment (assuming that the Boltzmann constant to be $k_{B}\equiv 1$). 

Consider the TCL$2$ equation in the interaction picture, 
\begin{equation}
\dot{\boldsymbol{\varrho}}_{\mathsf{S}}(\tau)=-\textstyle{\int_{0}^{\tau}}ds\, \mathrm{Tr}_{\mathsf{B}}\big[\boldsymbol{H}_{\mathrm{I}}(\tau),[\boldsymbol{H}_{\mathrm{I}}(s), \boldsymbol{\varrho}_{\mathsf{S}}(\tau) \otimes \boldsymbol{\varrho}_{\mathsf{B}}(0)]\big],
\label{TCl2_2-equation-integralform_sm}
\end{equation}
where
\begin{align}
\boldsymbol{H}_{\mathrm{I}}(\tau)=\Omega \big(\sigma_{-}a^{\dagger}  e^{-{i}(\omega_{0}-\omega_{\mathrm{c}})\tau}+ \sigma_{+} a e^{i(\omega_{0}-\omega_{\mathrm{c}})\tau}\big).
\end{align}

\begin{figure*}[tp]
\includegraphics[width=5.6cm]{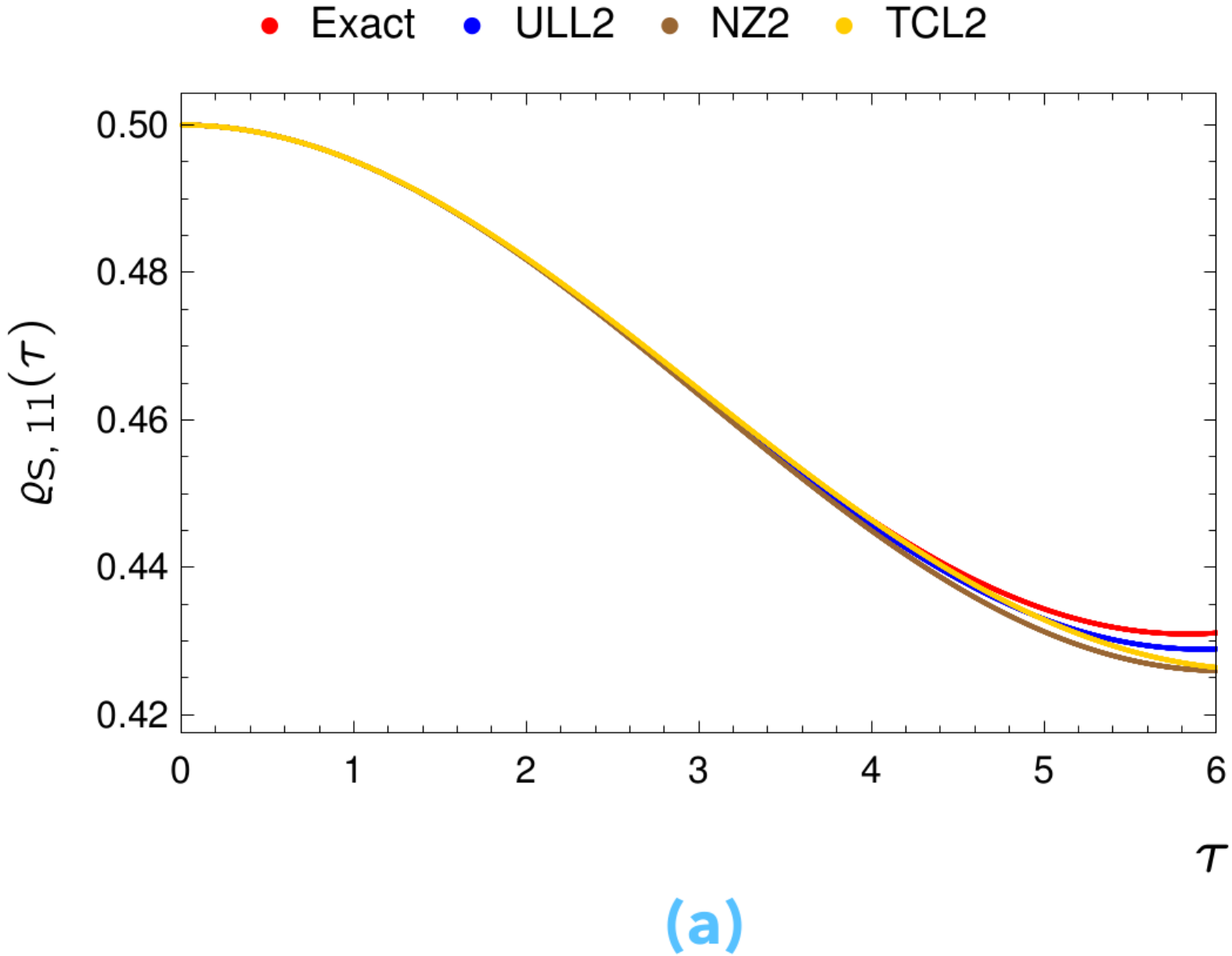} \hskip4mm 
\includegraphics[width=5.6cm]{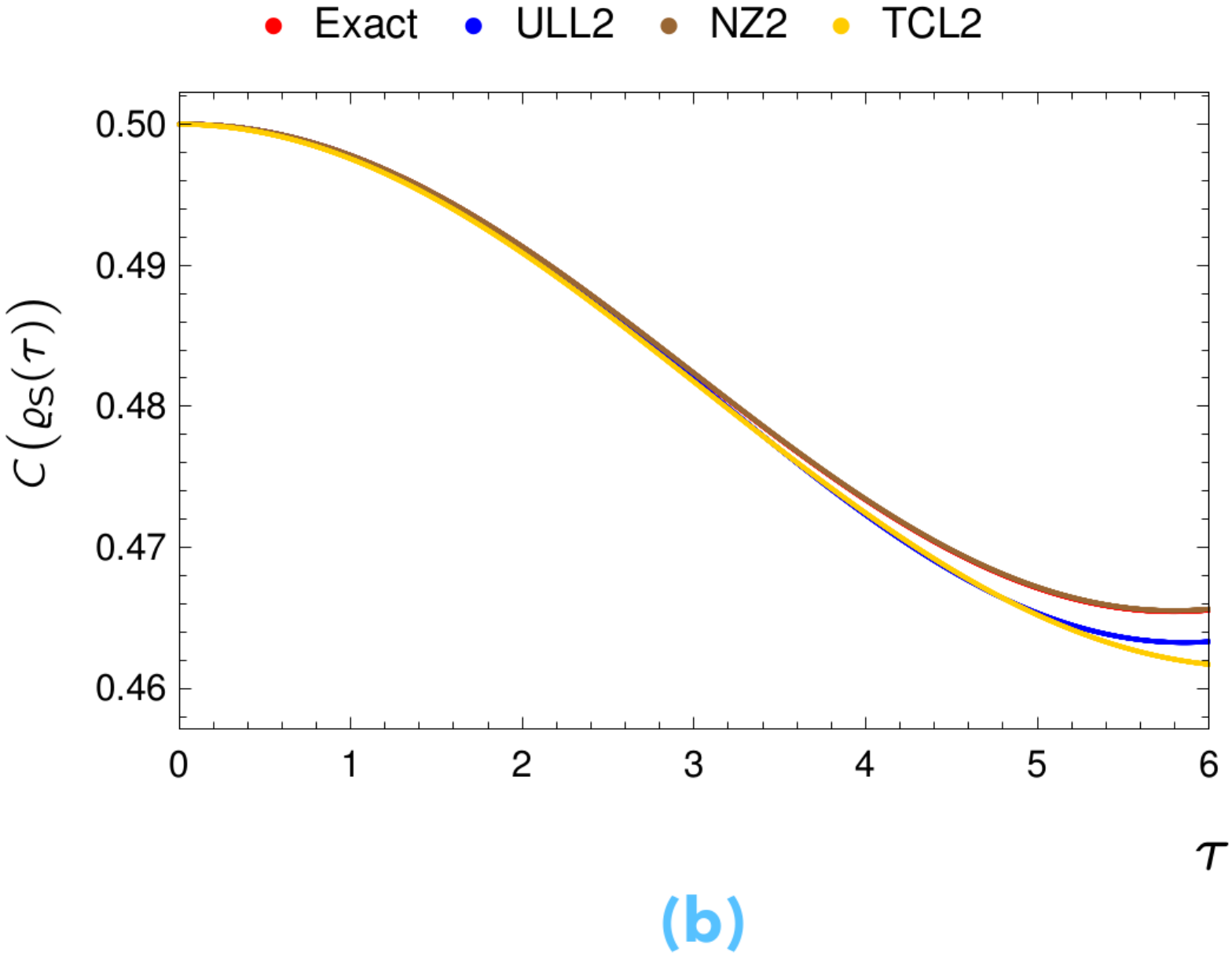} \hskip4mm 
\includegraphics[width=5.6cm]{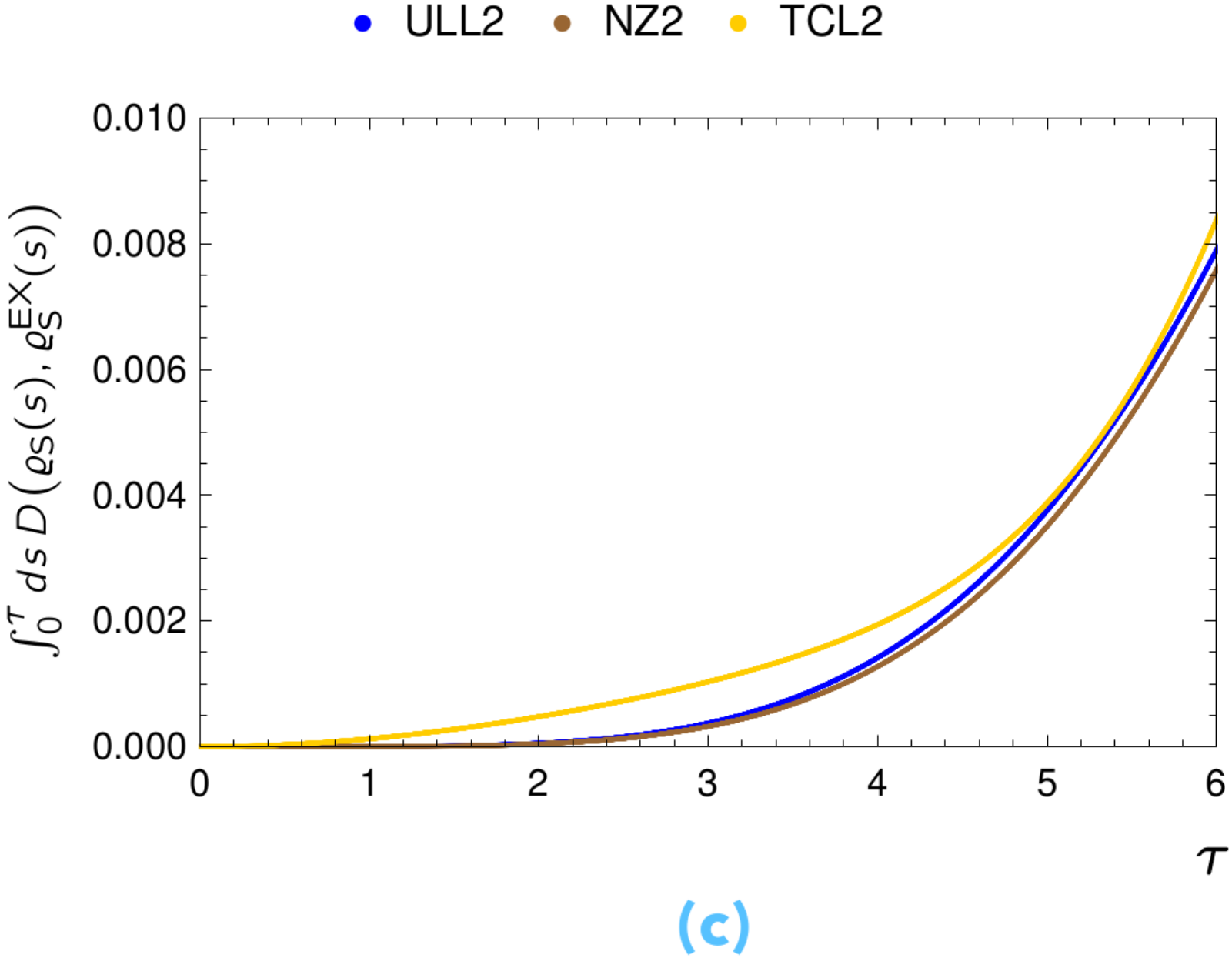}
\caption{Jaynes-Cummings model. (\textbf{\textsf{a}}) Dynamics of the excited-state population $\varrho_{\mathsf{S},11}$ and (\textbf{\textsf{b}}) coherence $C(\varrho_{\mathsf{S}})$ \textit{vs.} time in the exact ($\equiv$ ULL), ULL$2$, NZ$2$, and TCL$2$ techniques. (\textbf{\textsf{c}}) Accumulative error in the approximated state of the {system}. In (\textbf{\textsf{b}}) the exact and NZ$2$ curves overlap. In all techniques except ULL$2$ a constant state is assumed for the environment as $\varrho_{\mathsf{B}}(\tau)=\varrho_{\mathsf{B}}(0)$. The values of the parameters are $N=128$, $\Omega/\omega_{0}=0.1$, $\omega_{\mathrm{c}}-\omega_{0}=0.5$, $\beta=5$, and $\delta \tau = 0.0005$, and the initial state of the system is $(|0\rangle+|1\rangle)/\sqrt{2}$. All quantities are in natural units where $\hbar\equiv k_{B} \equiv 1$.}
\label{JC_model1}
\end{figure*}

\begin{figure*}[bp]
\includegraphics[width=5.6cm]{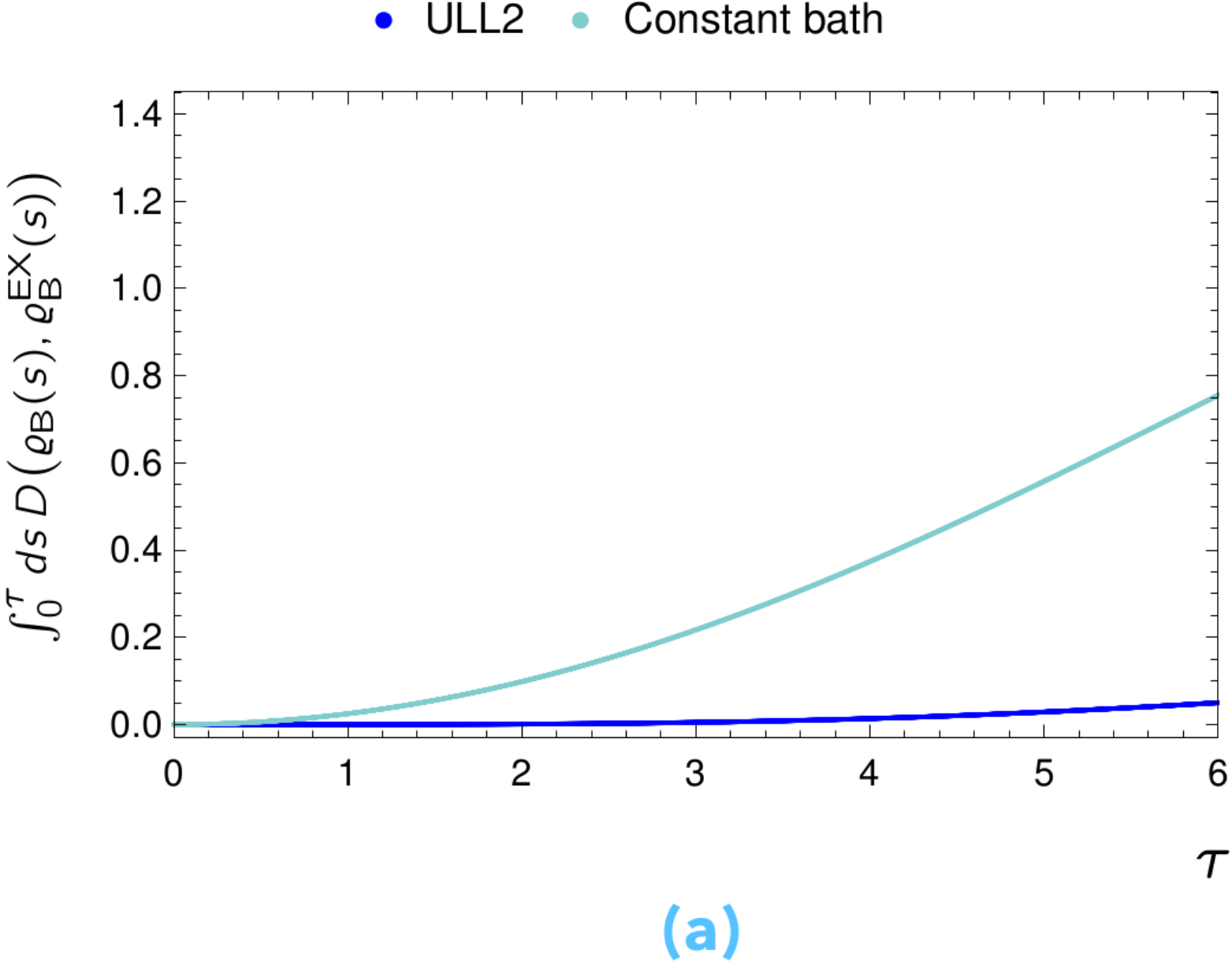} \hskip4mm
\includegraphics[width=5.6cm]{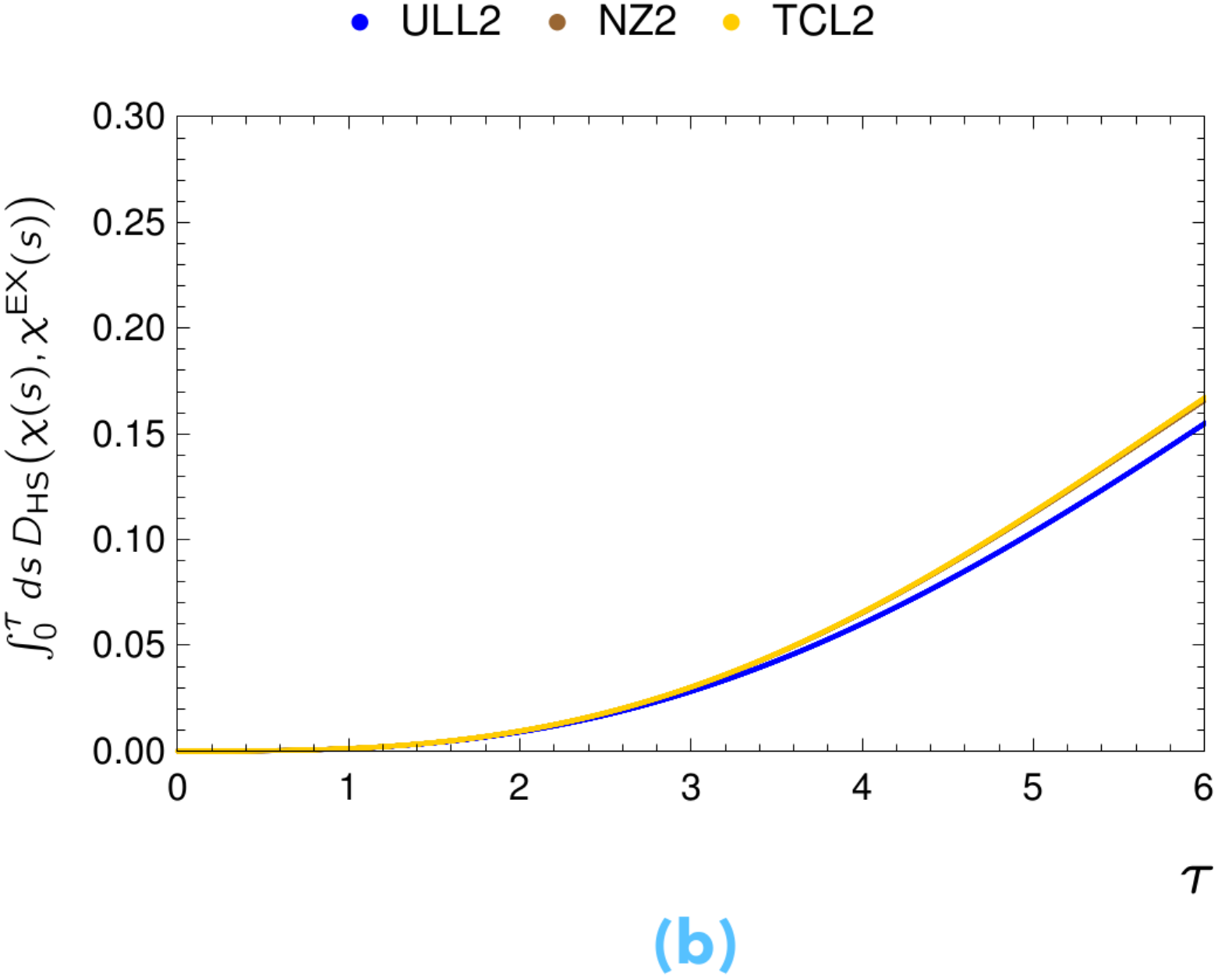}\hskip4mm
\includegraphics[width=5.6cm]{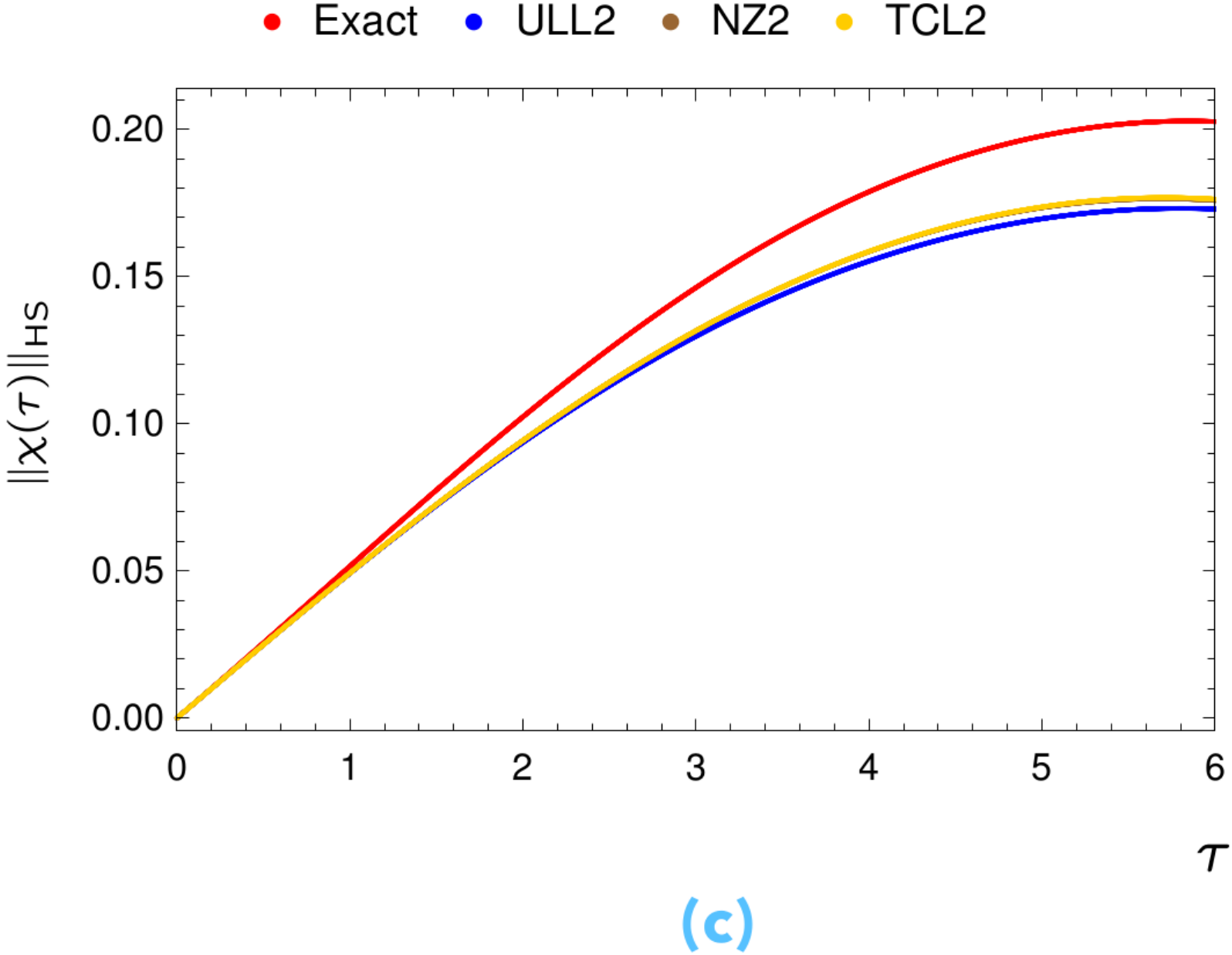}
\caption{Jaynes-Cummings model. (\textbf{\textsf{a}}) Error in capturing the dynamics of the {environment} state, (\textbf{\textsf{b}}) error in capturing the dynamics of the correlation operator, and (\textbf{\textsf{c}}) norm of the correlation operator vs. time in the exact dynamics (``Exact'' \& ``EX'' ), ULL$2$, NZ$2$, and TCL$2$ techniques. In (\textbf{\textsf{b}}) and (\textbf{\textsf{c}}) the NZ$2$ and TCL$2$ curves overlap. The values of the parameters are $N=128$, $\Omega/\omega_{0}=0.1$, $\omega_{\mathrm{c}}-\omega_{0}=0.5$, $\beta=5$, and $\delta \tau = 0.0005$, and the initial state of the system is $(|0\rangle+|1\rangle)/\sqrt{2}$. All quantities are in natural units where $\hbar \equiv k_{B} \equiv 1$.}
\label{Jc_model2}
\end{figure*}

We need to evaluate each term in the integral (\ref{TCl2_2-equation-integralform_sm}). For the first term we have
\begin{align}
\mathrm{Tr}_{\mathsf{B}} [\boldsymbol{H}_{\mathrm{I}}(\tau)\boldsymbol{H}_{\mathrm{I}}(s) \boldsymbol{\varrho}_{\mathsf{S}}(\tau ) \otimes \boldsymbol{\varrho}_{\mathsf{B}}(0)]=&\Omega^{2} \big(\sigma_{-}\sigma_{-} \boldsymbol{\varrho}_{\mathsf{S}}(\tau) e^{- i(\omega_{0}-\omega_{\mathrm{c}})(\tau+s)} \mathrm{Tr}_{\mathsf{B}}[a^{\dag} a^{\dag} \boldsymbol{\varrho}_{\mathsf{B}}(0) ]+ \sigma_{+}\sigma_{+}  \boldsymbol{\varrho}_{\mathsf{S}}(\tau)e^{ i(\omega_{0}-\omega_{\mathrm{c}})(\tau+s)} \mathrm{Tr}_{\mathsf{B}}[a^{2} \boldsymbol{\varrho}_{\mathsf{B}}(0) ]\nonumber\\&+ \sigma_{-}\sigma_{+} \boldsymbol{\varrho}_{\mathsf{S}}(\tau) e^{- i(\omega_{0}-\omega_{\mathrm{c}})(\tau-s)} \mathrm{Tr}_{\mathsf{B}}[a^{\dag} a \boldsymbol{\varrho}_{\mathsf{B}}(0) ]+ \sigma_{+}\sigma_{-}  \boldsymbol{\varrho}_{\mathsf{S}}(\tau)e^{ i(\omega_{0}-\omega_{\mathrm{c}})(\tau-s)} \mathrm{Tr}_{\mathsf{B}}[a a^{\dag} \boldsymbol{\varrho}_{\mathsf{B}}(0) ]\big).
\end{align}
Note that $\mathrm{Tr}_{\mathsf{B}}[a a^{\dag} \boldsymbol{\varrho}_{\mathsf{B}}(0)]=1+\langle n_{\mathsf{B}} \rangle$, where $\langle n_{\mathsf{B}} \rangle=\mathrm{Tr}_{\mathsf{B}}[ a^{\dag} a \boldsymbol{\varrho}_{\mathsf{B}}(0) ]$, and $\mathrm{Tr}_{\mathsf{B}}[a a \boldsymbol{\varrho}_{\mathsf{B}}(0) ]=\mathrm{Tr}_{\mathsf{B}}[a^{\dag}  a^{\dag} \boldsymbol{\varrho}_{\mathsf{B}}(0) ]=0$ for the thermal state of the environment. Thus,
\begin{align}
\mathrm{Tr}_{\mathsf{B}} [\boldsymbol{H}_{\mathrm{I}}(\tau)\boldsymbol{H}_{\mathrm{I}}(s) \boldsymbol{\varrho}_{\mathsf{S}}(\tau ) \otimes \boldsymbol{\varrho}_{\mathsf{B}}(0)]=&\Omega^{2} \sigma_{-}\sigma_{+}  \boldsymbol{\varrho}_{\mathsf{S}}(\tau)e^{- i(\omega_{0}-\omega_{\mathrm{c}})(\tau-s)} \langle n_{\mathsf{B}} \rangle+\Omega^{2} \sigma_{+}\sigma_{-} \boldsymbol{\varrho}_{\mathsf{S}}(\tau) e^{ i(\omega_{0}-\omega_{\mathrm{c}})(\tau-s)} (1+\langle n_{\mathsf{B}}\rangle).
\end{align}
Similarly, we obtain the other terms as 
\begin{align}
&\mathrm{Tr}_{\mathsf{B}} [\boldsymbol{H}_{\mathrm{I}}(\tau)\boldsymbol{\varrho}_{\mathsf{S}}(\tau ) \otimes \boldsymbol{\varrho}_{\mathsf{B}}(0) \boldsymbol{H}_{\mathrm{I}}(s) ]=\Omega^{2} \sigma_{-} \boldsymbol{\varrho}_{\mathsf{S}}(\tau)\sigma_{+} e^{- i(\omega_{0}-\omega_{\mathrm{c}})(\tau-s)} (1+\langle n_{\mathsf{B}} \rangle)+\Omega^{2} \sigma_{+} \boldsymbol{\varrho}_{\mathsf{S}}(\tau)\sigma_{-} e^{ i(\omega_{0}-\omega_{\mathrm{c}})(\tau-s)} \langle n_{\mathsf{B}}\rangle,\\&\mathrm{Tr}_{\mathsf{B}} [\boldsymbol{H}_{\mathrm{I}}(s)\boldsymbol{\varrho}_{\mathsf{S}}(\tau ) \otimes \boldsymbol{\varrho}_{\mathsf{B}}(0) \boldsymbol{H}_{\mathrm{I}}(\tau) ]=\Omega^{2} \sigma_{-} \boldsymbol{\varrho}_{\mathsf{S}}(\tau)\sigma_{+} e^{ i(\omega_{0}-\omega_{\mathrm{c}})(\tau-s)} (1+\langle n_{\mathsf{B}} \rangle)+\Omega^{2} \sigma_{+} \boldsymbol{\varrho}_{\mathsf{S}}(\tau)\sigma_{-} e^{- i(\omega_{0}-\omega_{\mathrm{c}})(\tau-s)} \langle n_{\mathsf{B}}\rangle,\\&   \mathrm{Tr}_{\mathsf{B}} [ \boldsymbol{\varrho}_{\mathsf{S}}(\tau ) \otimes \boldsymbol{\varrho}_{\mathsf{B}}(0)\boldsymbol{H}_{\mathrm{I}}(s)\boldsymbol{H}_{\mathrm{I}}(\tau)]=\Omega^{2} \boldsymbol{\varrho}_{\mathsf{S}}(\tau)\sigma_{-}\sigma_{+}  e^{ i(\omega_{0}-\omega_{\mathrm{c}})(\tau-s)} \langle n_{\mathsf{B}} \rangle+\Omega^{2} \boldsymbol{\varrho}_{\mathsf{S}}(\tau) \sigma_{+}\sigma_{-}  e^{ -i(\omega_{0}-\omega_{\mathrm{c}})(\tau-s)} (1+\langle n_{\mathsf{B}}\rangle) .
\end{align}
Putting the above terms together yields
\begin{equation}
\begin{split}
\dot{\boldsymbol{\varrho}}_{\mathsf{S}}(\tau)=& -i\epsilon^{{\textsc{tcl}\scriptscriptstyle{2}}}[(1+\langle n_{\mathsf{B}}\rangle)  \sigma_{+} \sigma_{-}-\langle n_{\mathsf{B}}\rangle\sigma_{-}\sigma_{+},\boldsymbol{\varrho}_{\mathsf{S}}(\tau) ]\\ 
&+(1+\langle n_{\mathsf{B}}\rangle) \gamma^{{\textsc{tcl}\scriptscriptstyle{2}}} [2\sigma_{-} \boldsymbol{\varrho}_{\mathsf{S}}(\tau)\sigma_{+}-\{\sigma_{+} \sigma_{-}, \boldsymbol{\varrho}_{\mathsf{S}}(\tau)\}]+\langle n_{\mathsf{B}}\rangle \gamma^{{\textsc{tcl}\scriptscriptstyle{2}}} [2\sigma_{+} \boldsymbol{\varrho}_{\mathsf{S}}(\tau)\sigma_{-}-\{\sigma_{-} \sigma_{+}, \boldsymbol{\varrho}_{\mathsf{S}}(\tau)\}],
\end{split}
\label{eq_tcl2_{j}Cmodel}
\end{equation}
where $\gamma^{{\textsc{tcl}\scriptscriptstyle{2}}} = \mathrm{Re}\big(\textstyle{\int_{0}^{\tau}} ds\, \Omega^{2}  e^{ i(\omega_{0}-\omega_{\mathrm{c}})(\tau-s)}\big)$ and $\epsilon^{{\textsc{tcl}\scriptscriptstyle{2}}} = \mathrm{Im}\big(\textstyle{\int_{0}^{\tau}} ds\, \Omega^{2} e^{ i(\omega_{0}-\omega_{\mathrm{c}})(\tau-s)}\big)$. The Redfield equation can be obtained by replacing $s\to\tau-s$ and setting the limit of integration to infinity in the TCL$2$ equation. Thus, the rate for the Redfield equation is 
\begin{equation}
\gamma^{{\textsc{r}}}=\mathrm{Re}\big(\textstyle{\int_{0}^{\infty}} ds\, \Omega^{2} e^{ i(\omega_{0}-\omega_{\mathrm{c}})s}\big) =\pi \Omega^{2} \,\delta(\omega_{0}-\omega_{\mathrm{c}}).
\end{equation}
Note that the dissipative part of the Redfield equation is zero for the case $\omega \neq \omega_{0}$. The Lindblad and the Redfield equations are equivalent for this example because $H_{\mathsf{S}}$ has only one energy gap. In Figs. \ref{JC_model1}, we show the dynamics of the density matrix {of the system} calculated with the NZ$2$, TCL$2$, ULL$2$, and exact techniques. The ULL$2$ and NZ$2$  solutions follows the actual dynamics and the coherence relatively better than TCL$2$, and hence the integrated distance is also relatively small for the ULL$2$ and NZ$2$ solutions. To evaluate how well the approximative techniques capture correlation, we also show the norm and the distance of the approximate correlation operator from the exact correlation in Figs. \ref{Jc_model2}. These results again demonstrate relative advantage of ULL$2$ over the correlation calculated with other techniques. 

%%%%%%%%%%%%%%%%%%%%%%%%%%%%%%%%%%%%%%%%%%%%%%%%%%%%%%%%%%%%%%%%
%%%%%%%%%%%%%%%%%%%%%%%%%%%%%%%%%%%%%%%%%%%%%%%%%%%%%%%%%%%%%%%%

%%%%%%%%%%%%%%%%%%%%%%%%%%%%%%%%%%%%%%%%%%%%%%%%%%%%%%%%%%%%%%%%
%%%%%%%%%%%%%%%%%%%%%%%%%%%%%%%%%%%%%%%%%%%%%%%%%%%%%%%%%%%%%%%%

%%%%%%%%%%%%%%%%%%%%%%%%%%%%%%%%%%%%%%%%%%%%%%%%%%%%%%%%%%%%%%%%
\end{document}